\documentclass[preprint,times]{elsarticle}
\pdfoutput = 1
\usepackage{amsmath}  
\usepackage{amssymb}
\usepackage{subfigure}
\usepackage{ecrc}
\usepackage{cuted}
\usepackage{rotating}
\usepackage{xcolor}
\usepackage{listings}
\usepackage{latexsym}
\usepackage{bm}
\usepackage{bbm}
\usepackage{floatrow}
\usepackage{breqn}
\usepackage{caption}
\newfloatcommand{capbtabbox}{table}[][\FBwidth]
\usepackage{tikz}
\definecolor{mygreen}{RGB}{28,172,0} 
\definecolor{mylilas}{RGB}{170,55,241}

\volume{00}

\firstpage{1}
\runauth{}

\begin{document}
\lstset{language=Matlab,%
    breaklines=true,%
    morekeywords={matlab2tikz},
    keywordstyle=\color{blue},%
    morekeywords=[2]{1}, keywordstyle=[2]{\color{black}},
    identifierstyle=\color{black},%
    stringstyle=\color{mylilas},
    commentstyle=\color{mygreen},%
    showstringspaces=false,
    numbers=left,%
    numberstyle={\tiny \color{black}},
    numbersep=9pt, 
    emph=[1]{for,end,break},emphstyle=[1]\color{red}, 
}

\begin{frontmatter}

\title{\textcolor{black}{Assessing size effects on the deformation of nanovoids in metallic materials}}

\author[]{J.~Hure\corref{cor1}}
\author[]{P.O. Barrioz}
\author[]{B. Tanguy}

\cortext[cor1]{Corresponding author: jeremy.hure@cea.fr, CEA Saclay, {91191} Gif-sur-Yvette, France}
\address[]{DEN-Service d'\'Etudes des Mat\'eriaux Irradi\'es, CEA, Universit\'e Paris-Saclay, F-91191, Gif-sur-Yvette, France}

\begin{abstract}
  An experimental methodology is developed to evaluate size effects in nanovoids deformation under macroscopic uniaxial stress loading conditions.
Quantitative evaluation of voids deformation as a function of voids size shows both a crystallographic effect, albeit small compared to the scatter, and no evidence of size effects for voids diameter larger than 10nm, while a slight effect is present for smaller voids. Critical assessment of the data in light of theoretical models indicates that these results may be compatible with the presence of a hardened layer at the void/matrix interface, which is illustrated through finite element simulations accounting for surface tension.
\end{abstract}

\begin{keyword}
  Porous material, Transmission electron microscopy, Dislocation, Size effects
\end{keyword}

\end{frontmatter}

Size effects related to the mechanical behavior of materials have been widely observed experimentally, \textit{e.g.}, the \textit{Hall-Petch effect} describing the dependence of the yield stress of a polycrystalline aggregate to the grain size \cite{hall,petch}, the \textit{Indentation Size effect} (ISE) in nanohardness measurements where an inverse relation between hardness and indentor depth is observed \cite{gane}, the \textit{Bending effect} of thin films showing a dependence of the angle-momentum relationship to the film's thickness \cite{stolken}. Explanations of these three examples of size effects have been proposed based on dislocation theory, by considering a \textit{plasticity lengthscale} related to the concept of Geometrically Necessary Dislocations (GNDs) \cite{ashby,smyshlyaev,nixgao}. Additional factors have also been shown to lead to size effects, such as dislocation scarcity implying a critical size related to dislocation density \cite{lilleodden}, or interfacial energy / surface stresses that may become relevant for small systems as the ratio between surface and volume decreases \cite{kramer}. These experimental observations have lead to the development of size-dependent plasticity models involving one or more lengthscales, most of them being referred to, broadly speaking, as strain-gradient plasticity models \cite{fleckhutchinson,fleckhutchinson2,forest}.\\
\indent
\textit{Void size effects} have been investigated theoretically and numerically in the context of porous materials modelling, relevant for ductile fracture through void growth to coalescence modelling \cite{benzergaleblond}. Molecular Dynamics (MD) simulations of nanoporous perfect crystals \cite{traiviratana} lead to the so-called \textit{smaller is stronger} effect where plasticity is mediated through dislocations nucleation at the void surface where surface tension can also play a role \cite{chang2013}. A \textit{larger is faster} effect is reported for void growth, explained through the density of dislocations sources in the bulk \cite{segurado2009}. Discrete Dislocation Dynamics (DDD) simulations showed the effect of dislocation scarcity on void growth \cite{chang2015}, and emphasized the role of GNDs to reduce void growth for smaller voids in the case of high stress triaxiality conditions, while limited effects are observed where strain-gradient are mostly absent \cite{borg2008}, \textit{e.g.}, under simple shear. These simulations are limited to low applied strain, so that conclusions are mostly relevant for early stages of porous materials deformation. Nevertheless, homogenized porous models have been developed that reproduce the \textit{smaller is stronger - larger is faster} size effects \cite{wen2005,monchiet2013b} in the context of ductile fracture modelling, mostly based on strain-gradient plasticity \cite{borg2008,niordson2008}.

  \begin{center}
\begin{figure}[H]
\includegraphics[height = 5.8cm]{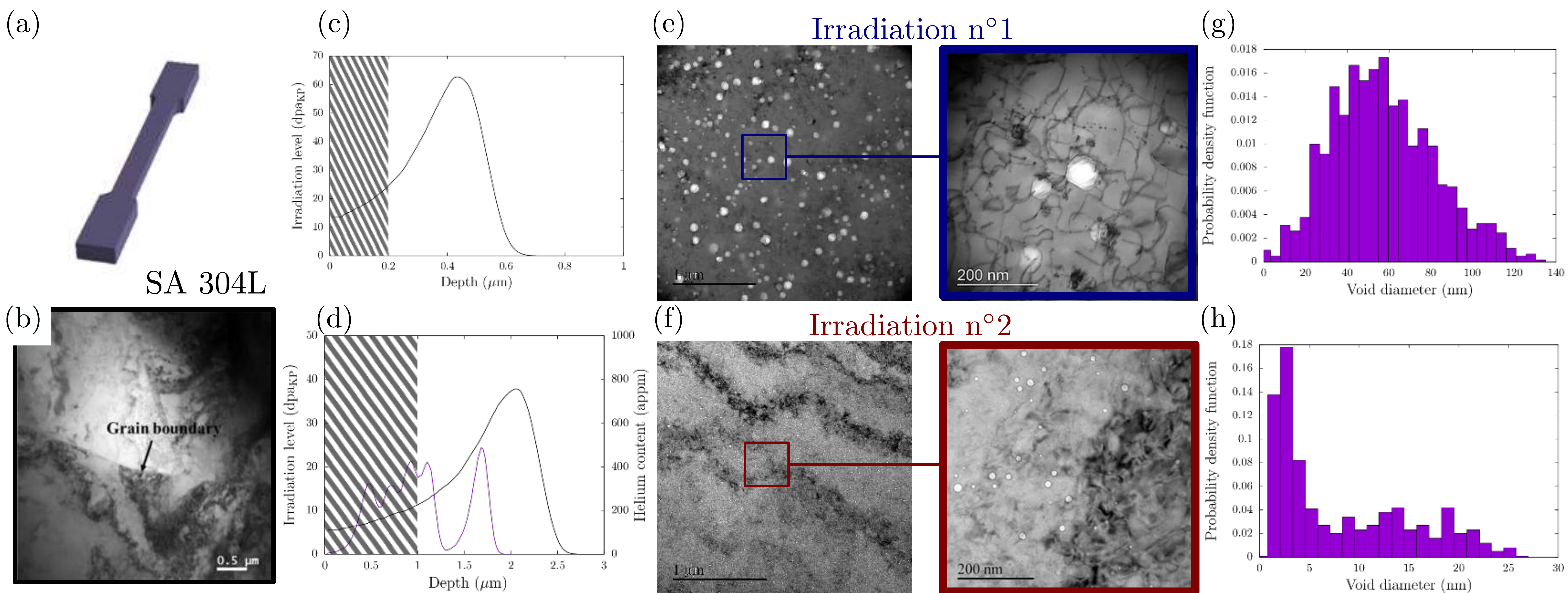}
\label{gra5}

  \caption{Experimental methodology to get nanoporous materials: (a,b) Tensile samples made of Solution Annealed (SA) 304L austenitic stainless steel are irradiated with heavy-ions under two different conditions (c,d), leading to a micrometric irradiated layer composed of nanovoids and dislocations. For each irradiation condition, ion depth and damage profile (calculated using SRIM software \cite{SRIM} with Kinchin-Pease approximation, and assuming a threshold displacement energy of $40$eV), typical TEM images after irradiation (e,f) and voids size distributions (g,h) are reported.}
\end{figure}
  \end{center}

Experimental assessment of size effects on voids deformation under mechanical loading or on strength of porous materials remains scarce, and, to the authors' knowledge, no experimental evidence of void size effects is available.
The case of nanoporous metallic materials is particularly interesting, as small voids embedded in a single crystal can show both size and crystallographic orientation effects. Analysis of fracture surfaces \cite{margolin2016,neustroev} or in-situ Transmission Electron Microscope tensile tests of nanoporous metallic materials \cite{ding,han} clearly indicate that nanovoid ($<100$nm) growth and coalescence are qualitatively similar to their larger counterparts, but no quantitative experimental data has been reported for nanovoid deformation, which is therefore the aim of this study.\\
\indent
A commercial-grade Solution-Annealed austenitic stainless steel 304L is used.
Tensile samples (Fig.~1a), made by electrical discharge machining technique followed by a mirror polishing (up to 0.05$\mu$m colloidal silica vibratory polishing), are irradiated with heavy-ions at the JANNuS accelerators facility \cite{jannus}. Under specific conditions, irradiation with high energy particles of metallic materials has been shown to lead to nanovoids \cite{cawthorne}.
Irradiation is thus used here as a tool to generate nanoporous materials. Two different irradiation conditions have been selected based on literature data to get different void sizes distributions, and the parameters are summarized in Tab.~1. Ion penetration depth and damage level have been calculated using SRIM software \cite{SRIM}, as shown in Fig.~1c,d, and the typical irradiation depth is on the order of 1 $\mu m$. The TEM observations were performed at the surface after removing $200\mathrm{nm}$ (resp. $1\mu m$) for Irradiation n$^{\circ}$1 (resp. n$^{\circ}$2), as shown on Fig.~1, in order to have a higher void density. Details about the material, sample preparation, irradiation conditions and TEM observations can be found in Supplementary Material.
\begin{center}
  \begin{table}
    \scalebox{0.75}{
  \begin{tabular}{ccccc}
    \hline
    \hline
 Irradiation & Ion & Temperature & Flux & Fluence \\
 & &  ($^{\circ}$C) & ($\mathrm{ions.cm^{-2}.s^{-1}}$) & ($\mathrm{ions.cm^{-2}}$) \\
 \hline
 \hline
1 &  C 0.5MeV  & 600 & $1.8\ 10^{13}$ & $4.0\ 10^{17}$\\
 \hline
2 &  He 1MeV  & 20  &$4.2\ 10^{11}$ & $1.0 \ 10^{16}$ \\
  &Fe 10MeV  & 600 &$4.6\ 10^{11}$ &$4.0 \ 10^{16}$ \\
 \hline
 \hline
  \end{tabular}
  }
  \caption{Irradiation conditions. Irradiation n$^{\circ}$2 was performed in two steps: Helium implantation at room temperature, followed by Fe-irradiation at high temperature.}
  \end{table}
\end{center}
\vspace{-0.8cm}

\indent
After irradiation, TEM analyses have been performed on the irradiated samples. Typical images are given in Fig.~1e,f, where the microstructure is composed mainly of voids and a high density of dislocations. For both irradiations, the dislocation density is estimated to be about $5.10^{14}\mathrm{m^{-2}}$ before straining. Voids appear spherical or slightly faceted for the larger ones as a result of the minimization of surface energy that depends on interface orientation \cite{wulff}. TEM images analysis allows to get void diameter distributions as shown on Fig.~1g,h. The two irradiations lead to very different voids size distributions, with mean diameter of about $50\mu$m and $5\mu$m, respectively, confirming the ability to use ion-irradiation as a tool to generate nanoporous materials by changing the irradiation parameters.
\begin{center}
\begin{figure}[H]
  \subfigure[]{\includegraphics[height = 3.5cm]{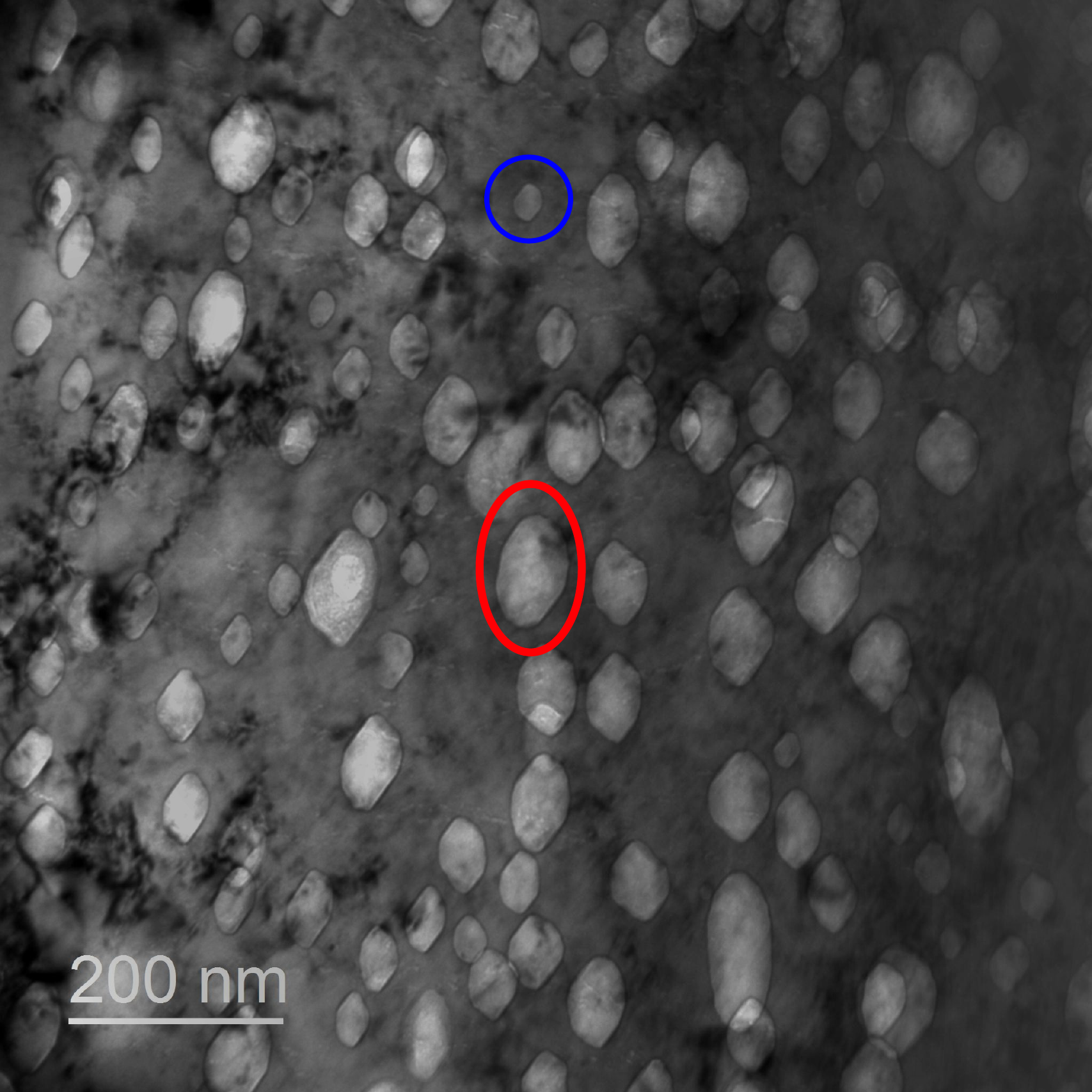}}
  \subfigure[]{\includegraphics[height = 3.5cm]{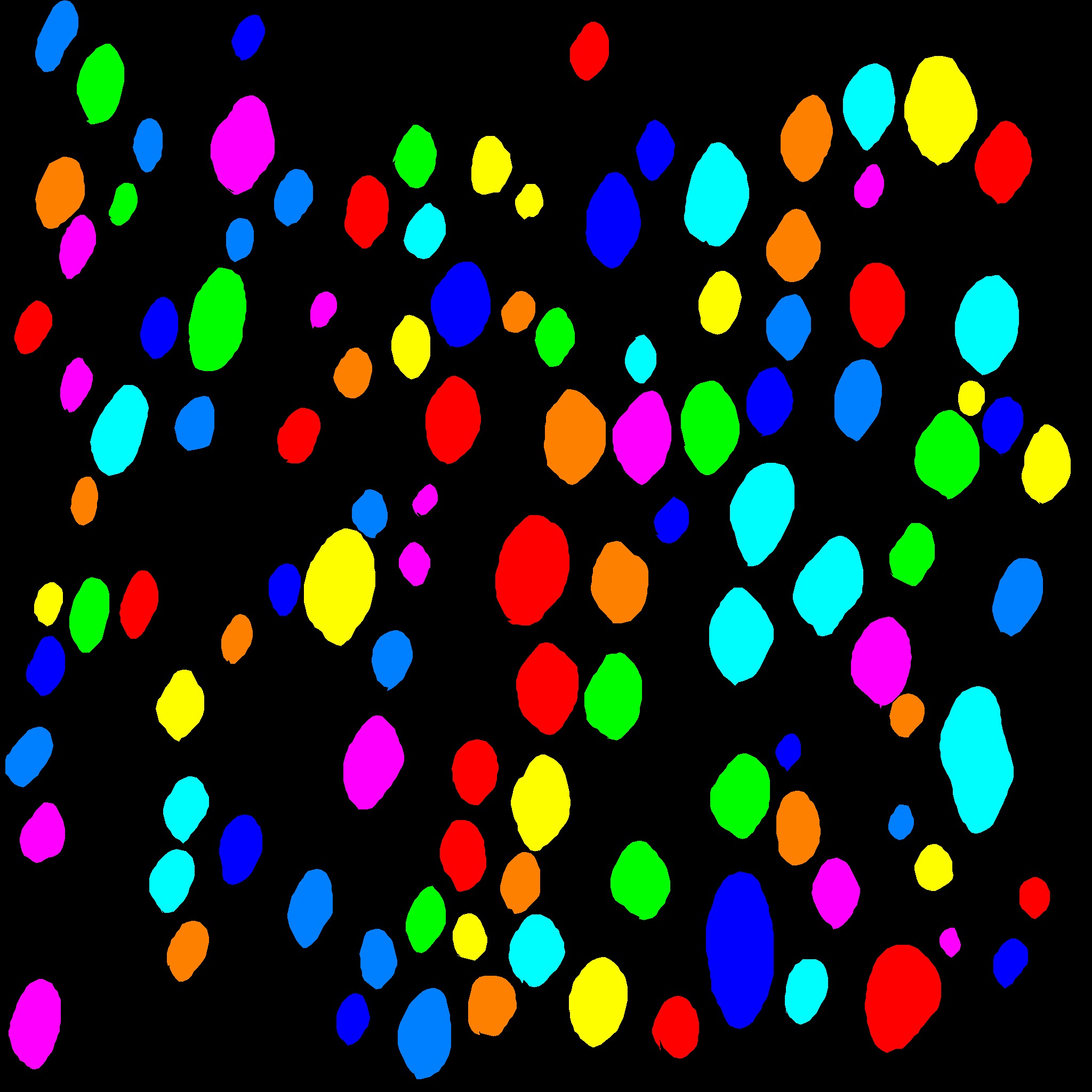}}
      \subfigure[]{\includegraphics[height = 3.5cm]{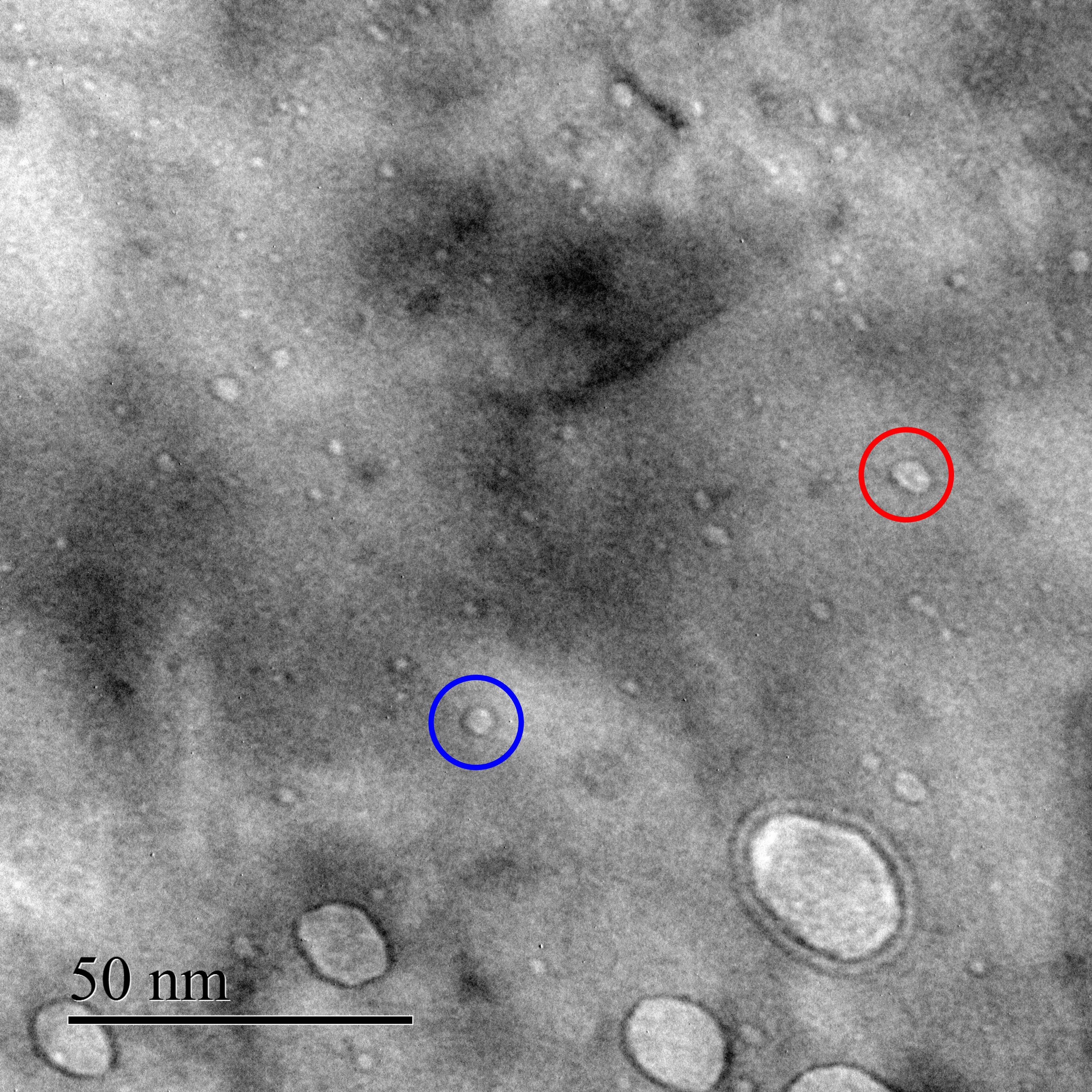}}
    \subfigure[]{\includegraphics[height = 3.5cm]{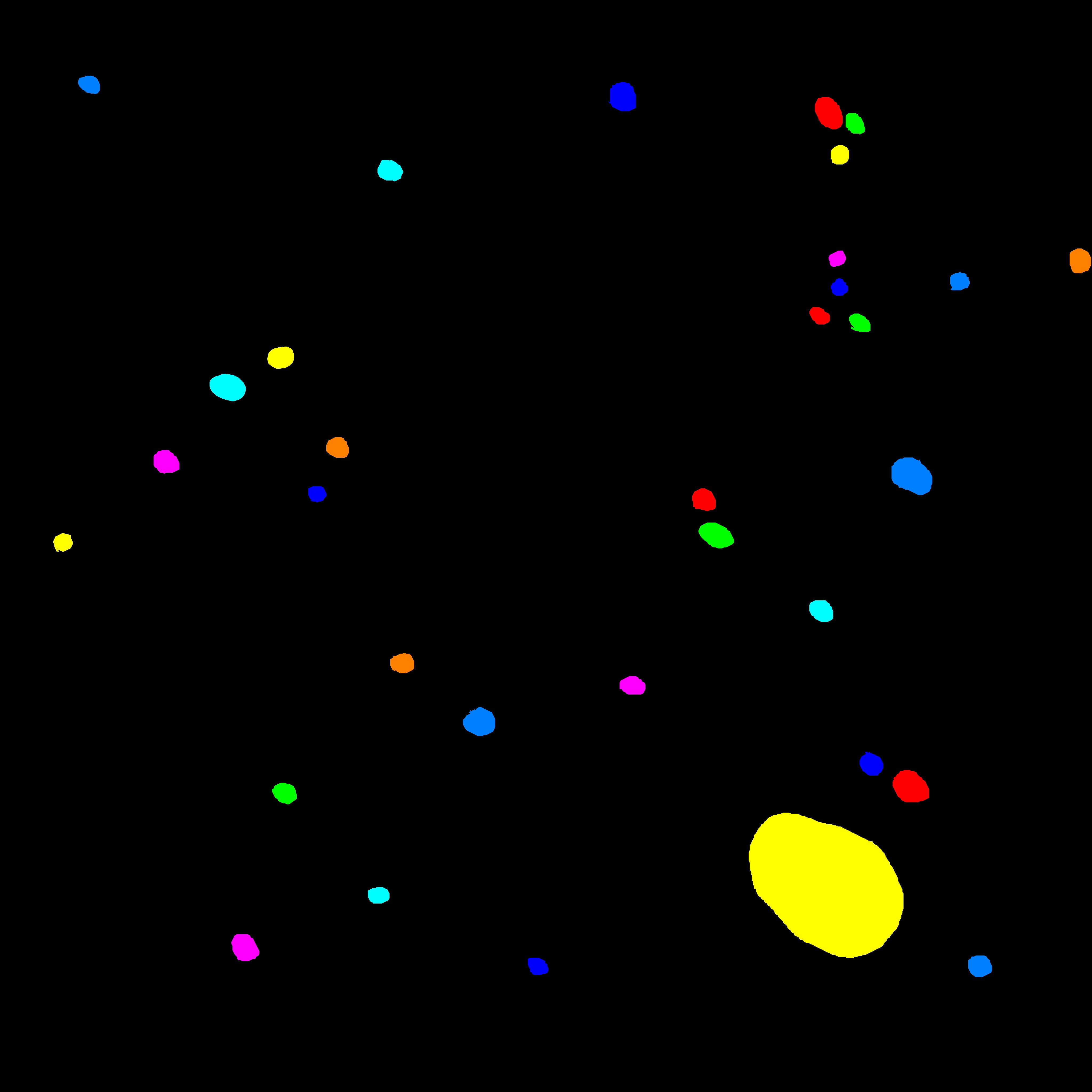}}
    \label{figdef}
    \caption{TEM observations of deformed nanovoids after tensile test at 300$^{\circ}$C up to 30\% strain, for (a) Irradiation n$^{\circ}1$ and (c) n$^{\circ}2$. Images (b) and (d) correspond to binarization of voids with an in-house software. Examples of highly deformed and mostly undeformed voids are circled in red and blue, respectively.}
\end{figure}
\end{center}
\vspace{-1cm}
As reported in the literature \cite{was}, pre-implantation of Helium allows nucleating numerous cavities, of smaller size that without pre-implantation, but potentially leading to an additional difference than void size regarding the two irradiations, due to Helium pressure. An estimate of Helium pressure in each cavity can be obtained using the following assumptions: all implanted He atoms are inside the cavities of same diameter in equivalent quantity, assuming ideal gas law, which leads to $p_{He} \sim 20\mathrm{MPa}$, which is small compared the Young-Laplace pressure generated by surface tension $\gamma \sim 1\mathrm{J.m^{-2}}$ \cite{crystallium2} of the void/matrix interface $p_{\gamma} = 2\gamma/R \sim 200\mathrm{MPa}$ for $R \sim 10\mathrm{nm}$. Therefore, in the following, the only difference between the two irradiations is assumed to be the voids size distributions. Irradiated tensile samples were tested at $300^{\circ}$C with a conventional tensile machine at a mean strain rate $\dot{\varepsilon} = 5.10^{-4}\mathrm{s^{-1}}$. The test temperature was selected in order to have a deformation mode based on dislocation glide, by avoiding potential twinning that may appear at room temperature \cite{bellefon}. TEM observations were finally performed on thin foils extracted from tested irradiated samples, at different locations at the surface of the tensile samples, thus corresponding to different crystallographic orientations.\\
\begin{center}
\begin{figure}[H]
  \subfigure[]{\includegraphics[height = 4.5cm]{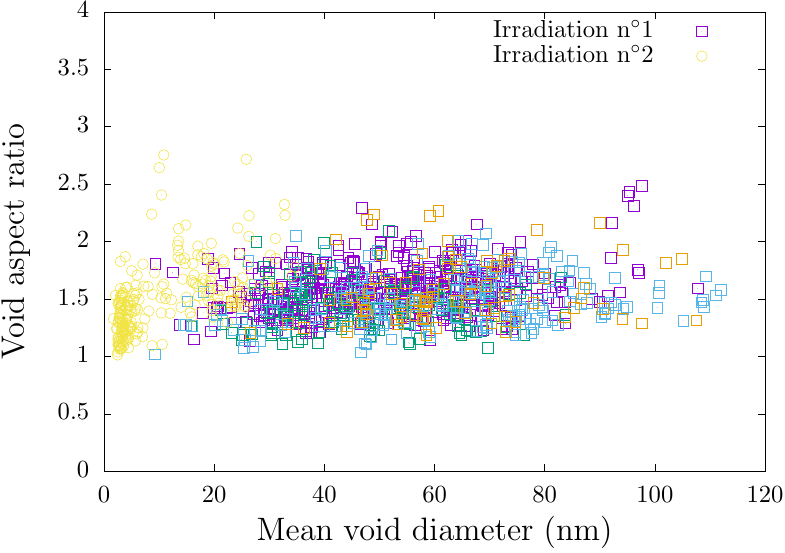}}
    \subfigure[]{\includegraphics[height = 4.5cm]{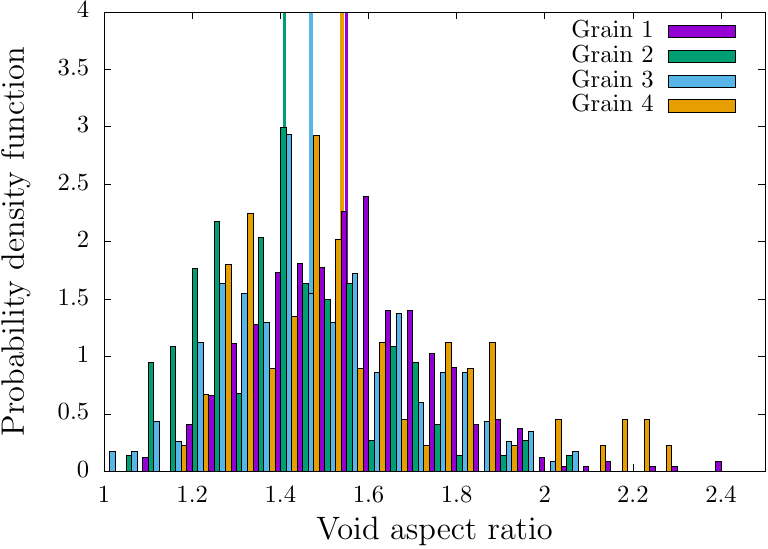}}
    \label{figres}
    \caption{(a) Void aspect ratio (defined using the axis $a$ and $b$ of an ellipse inscribed into the voids) as a function of mean void diameter $\sqrt{ab}$. For irradiation n$^{\circ}$1, colors correspond to observations in different grains. (b) Probability density function of void aspect ratio in different grains, for irradiation n$^{\circ}$1: vertical lines correspond to mean values for each grain.}
\end{figure}
\end{center}
\vspace{-1cm}
\indent
Typical results are shown on Fig.~2 where deformed nanovoids are shown \textcolor{black}{after tensile test at 300$^{\circ}$C up to 30\% strain.} A first qualitative observation is that deformation is rather homogeneous at the scale of the voids, \textit{i.e.}, no steps resulting from the inherent discrete nature of dislocations glide are observed on voids surfaces (at least for larger voids) and nanovoids shapes are reminiscent of typical observations of large scale voids under uniaxial stress loading conditions \cite{wecktoda}. Secondly, noticeable differences appear regarding void deformation on Fig.~2a where some voids are clearly elongated along the tensile direction whereas others are almost undeformed. Voids deformation is thus quantified by the aspect ratio of the ellipse of axis $a$ and $b$ inscribed into the voids, and plotted as a function of a function of the mean void diameter defined as $D = \sqrt{ab}$ in Fig.~3. As TEM images correspond to projections of voids on the observation plane, such definition of void aspect ratio implicitly assumes that voids are spheroidal with respect to the tensile axis, which might not be the case due to the anisotropy of single crystals. This point will be discussed later on based on numerical results. As shown in Fig.~3a, both irradiations lead to consistent results on their overlapping voids size range. A large scatter of the void aspect ratio is observed for a mean void diameter (Fig.~3a), similarly to what have already been observed at higher scales \cite{lecarmemaire}. As voids in different crystallographic orientations subjected to the same loading conditions deform differently \cite{yerra}, a crystallographic effect might be at the origin of such scatter. Although an effect of crystallography is effectively observed regarding the average aspect ratio, as shown on Fig.~3b where observations in different grains are reported, the effect is small compared to the scatter observed in each grain.

Another potential source of scatter is that, whereas uniaxial stress loading conditions have been applied to the polycrystalline tensile samples up to a fixed macroscopic strain level, local stress/strains conditions at the single crystal scale might differ significantly. As shown in Supplementary material, deviations from uniaxial stress conditions are observed on polycrystalline aggregates, as well as locally higher (or lower) strain level. However, these heterogeneities are notably lower at the free surface where TEM observations have been performed, and appear at characteristic scales of the order of the grain size, thus can not explain the observed scatter at the $\mu$-scale (Fig.~3a) which is thus assumed to come from dislocations substructures at scales below the resolution of the TEM observations of this study. Finally, no evidence of size effects regarding the evolution of void aspect ratio as a function of mean void size is observed down to very low scales (about 10nm diameter). Below 10nm, although TEM observations become delicate (as blurriness leads to more apparent spherical voids), a slight size effect may be present, leading to lower mean aspect ratio. The mean aspect ratio can be understood as follows: assuming that strain is homogeneous down to void scale, any length $l$ along the tensile direction becomes $\varepsilon l$, while along the transverse direction $[-\varepsilon/2] l$ due to volume conservation at the macroscale. Thus, void aspect ratio is equal to $[R + \varepsilon R][R - (\varepsilon/2) R] \approx 1 + [3/2]\varepsilon = 1.45$, as $\varepsilon = 30\%$, which is in good agreement with the experimental data for larger voids (Fig.~3).
Such simplified model does not account for crystallographic effects, which will be discussed later based on numerical simulations.

The different physical mechanism proposed in the literature to explain (void) size effects are now critically assessed with respect to these experimental data. Initial dislocation density has been estimated to be about $\rho \sim 5.10^{14}\mathrm{m^{-2}}$, and increases drastically during the tensile test. A typical associated \textit{plasticity lengthscale} is $\rho^{-1/2} \lesssim 10\mathrm{nm}$, below most voids sizes considered in this study. This implies that plasticity can appear homogeneous down to the scale of the voids, explaining the smooth void shapes (in agreement with results from \cite{ding}), and also that dislocation scarcity \cite{chang2015} that may conduct to void size effect is not relevant for that particular conditions (consistently with recent observations on irradiated micropillars \cite{paccou}). More generally, even for lower initial dislocation densities, the strong increase due to plastic deformation - with typically saturation values reported as $\rho \sim 10^{15} - 10^{16} \mathrm{m^{-2}}$ \cite{fpz} - leads to assume that void size effect related to dislocation scarcity is relevant only for very low applied strain.
The effect of Geometrically Necessary Dislocations (GNDs) is expected to be relevant when their required density is large compared to the density of Statistically Stored Dislocations (SSD). GNDs arise due to strain gradients $\nabla \varepsilon$ \cite{ashby}, and the associated density scales as $\rho_{GND} \sim \nabla \varepsilon / b$, where $b$ is the magnitude of the Burgers vector. A simple estimate of the typical GNDs density can be made based on simplifying assumptions. Using Rice-Tracey void growth model \cite{ricetracey} for isotropic material modified by Huang \cite{huang}, local strain rate at the void scale is $\dot{\varepsilon}_V = \alpha T^{1/4} \exp{([3T/2])} \dot{\varepsilon}$, with $T$ the stress triaxiality and $\alpha$ a numerical prefactor, leading to $\dot{\varepsilon}_V \approx \beta \dot{\varepsilon}$ ($\beta = 1.25 \alpha$) for uniaxial stress loading conditions $T=1/3$. Strain gradient, on a typical size corresponding to void radius, can thus be estimated as $\nabla \epsilon \approx [\varepsilon - \varepsilon_V]/R \approx [1 - \beta] \varepsilon/R$. The corresponding GNDs density, taking the theoretical value of $\alpha = 0.427$ \cite{huang}, is $\rho_{GND} \approx [0.4 \varepsilon]/[bR] \approx 5.10^{16} \mathrm{m^{-2}}$ (resp. $5.10^{15} \mathrm{m^{-2}}$) for void size $R=10\mathrm{nm}$ (resp. $R=100\mathrm{nm}$). These estimates are on the same order as the dislocation density inferred from TEM observations. However, no clear sign of size effects has been observed on that range of void radii (Fig.~3a), which may indicate that the required density of GNDs has to be at least one order of magnitude higher than SSDs density in order to observe a size effect. In addition, experimental evaluations of the numerical prefactor $\alpha$ \cite{lecarmemaire} lead to values closer to unity, thus to very small required GNDs density, hence no size effects. The last theoretical argument proposed in the literature is surface tension effect due to the presence of a void/matrix interface \cite{dormieux2010}.
\begin{center}
\begin{figure}[H]
 \includegraphics[height = 4.5cm]{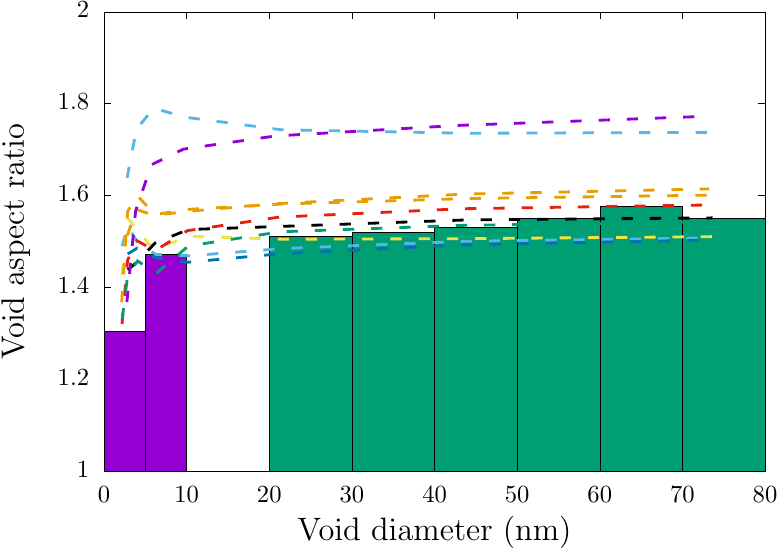}
    \label{figsim}
    \caption{Evolution of the mean void aspect ratio as a function of the mean void diameter. Experimental (boxes) and numerical (dashed lines with colors corresponding to different crystallographic orientations) results}
    \label{gra8}
\end{figure}
\end{center}
\indent
In order to get some insights about this effect, as well as to evaluate the effect of crystallographic orientation, finite element simulations have been performed on porous unit cell under periodic boundary conditions and macroscopic uniaxial tension. FCC crystal plasticity constitutive equations calibrated for 304L stainless steel accounting for the 12 slip systems are used, as well as a modelling of surface tension of magnitude $1\mathrm{J.m^{2}}$. Different random crystallographic orientations are used for the material (see Supplementary Material).\\

 Numerical results (Fig.~\ref{gra8}) show, for large voids, an absence of void size effects and a significant crystallographic effect. The average void aspect ratio (based on the different orientations) is close to the experimental value. For smaller void size, a decrease of the void aspect ratio is observed, limited to very small voids ($<10$nm) and dependent on crystallographic orientation. These observations show that, in addition to a potential GNDs effects discussed earlier, the effect of surface tension is a potential candidate to explain the size effect, albeit restricted to very small voids, affecting plastic flow under (or even without) mechanical loading. High-resolution TEM observations around the smallest voids are necessary to assess experimentally the potential physical mechanism, by looking at the dislocation density or if possible to assess the mechanical strains/stresses.\\

\indent
In summary, this study provides quantitative data regarding nanovoids deformation under mechanical loading, these data being also relevant for nuclear applications where similar nanoporous materials can be created under irradiation \cite{margolin2016}. Experimental conditions are relevant for low stress triaxiality conditions and high level of dislocation density (high strain), where the three main observations are that voids deform rather homogeneously down to very small scales, a large scatter regarding voids deformation at small scales - which is not accounted in any model - and a limited size effect for very small voids manifesting itself by a reduced void deformation. This size effect is limited to very small voids ($<10$nm), and a critical discussion with respect to the literature dealing with size effects and numerical simulations shows that the effects of GNDs and surface tension might be responsible for these effects. For the conditions used in this study, these results indicate that continuum mechanics modelling of plasticity can still hold at the nanometric scale, and that the scatter observed (similarly to what can be observed from X-ray tomography experiments) deserves attention regarding modelling of porous materials. \textcolor{black}{Although no size effect on nanovoid deformation has been observed down to 10nm void diameter, these results may not be in contradiction with simulations results based on strain gradient plasticity that predict size effects for voids typically smaller than one micron, as the applied stress triaxiality applied is usually much higher, therefore promoting strain gradients. The key point is that no systematic statement should be made regarding the length scale below which size effects are expected, as it strongly depends on both mechanical loading and microstructural features.} Experimental evidence of physical mechanism behind the potential size effect are still required - through high-resolution TEM observations and measurements of strains around small voids - and should be the object of future studies.\\

\noindent
\textbf{Acknowledgement}: The authors would like to thank JANNuS team for performing the ion irradiations. This research did not receive any specific grant from funding agencies in the public, commercial or not-for-profit sectors.

\newpage

\begin{center}
  {\Large Supplementary Material}
  \end{center}
\noindent
\textbf{Experiments: Materials and Methods}\\

\begin{figure}[H]
\centering
\subfigure[]{\includegraphics[height = 4.5cm]{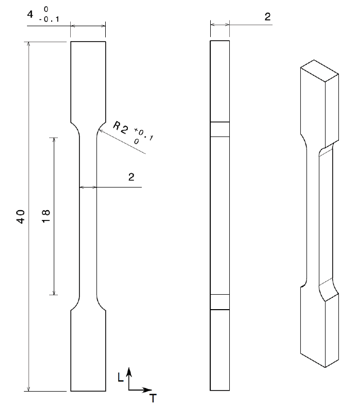}}
\hspace{0.5cm}
\subfigure[]{\includegraphics[height = 4.5cm]{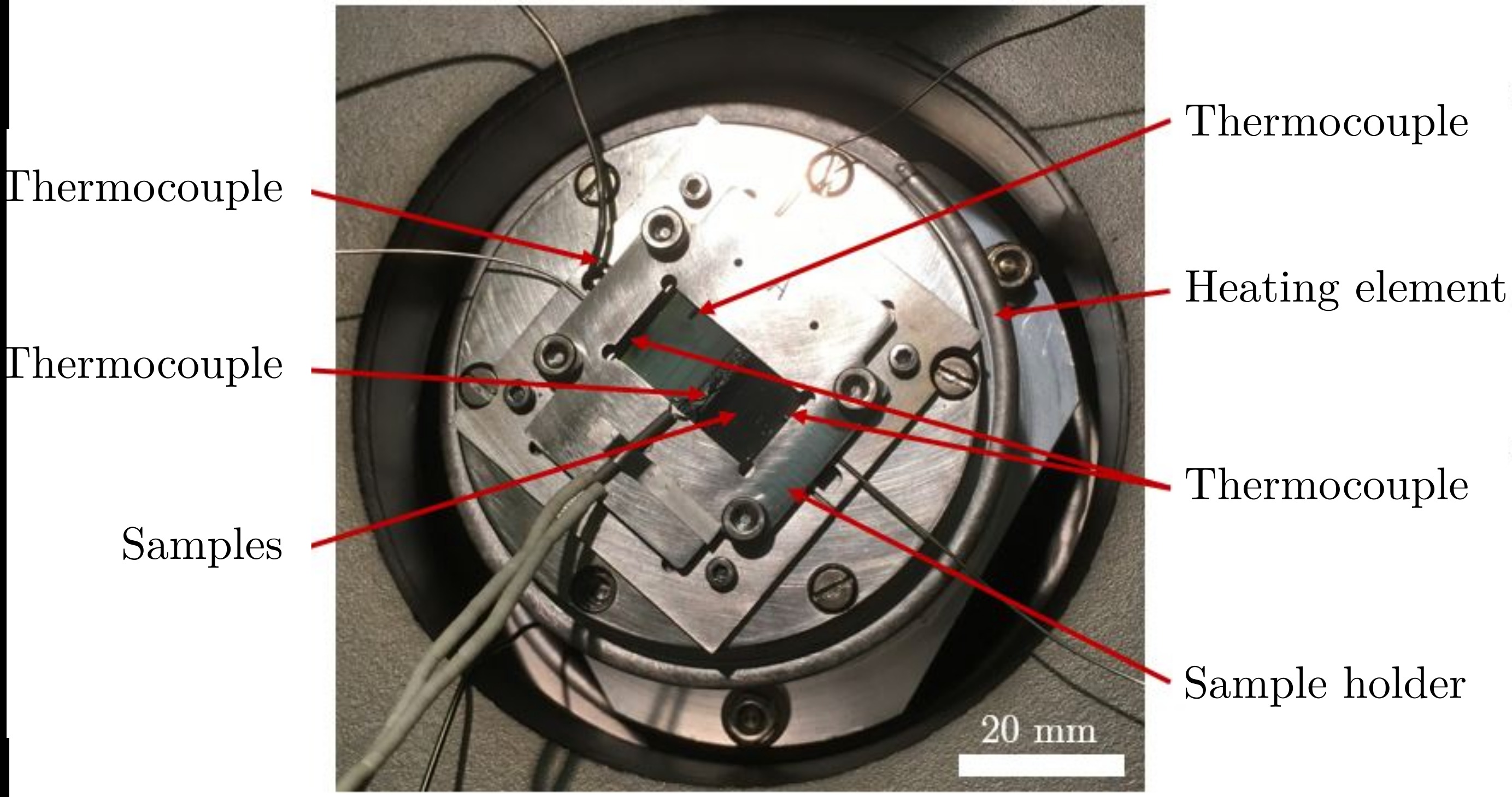}}
\caption{(a) Geometry of the samples (dimensions in mm) (b) Irradiation setup}
\label{sup1}
\end{figure}

In order to get nanoporous materials, commercial grade AISI 304L austenitic stainless steel samples have been irradiated with heavy ions. The chemical composition of the material is Fe - 18.75Cr - 8.55Ni - 0.02Mo - 0.45Si - 1.65Mn - 0.012C - 0.01P - 0.002S (wt \%) used in the Solution-Annealed (SA) condition (1050$^{\circ}$C/30min followed by water quenching). The mean grain size is about $30\mu$m. The microstructure is mainly Face-Centered-Cubic (FCC) austenite. The geometry of the irradiated samples obtained through electro-spark machining are shown on Fig.~\ref{sup1}a. Mechanical polishing has been performed on one side of the samples to remove surface roughness and hardened layer due to machining, down to 1$\mu$m diamond paste followed by vibratory polishing using 0.05$\mu$m colloidal silica for 10h. Such polishing procedure, already used in \cite{gupta2016}, leads to a surface free of defects (dislocations, nanograins). Irradiations have been performed at JANNuS Saclay irradiation facility \cite{jannus}. The irradiation setup is shown on Fig.~\ref{sup1}b: the sample holder is in contact with a heating element, and thermocouples in contact with the specimens are used to control temperature. For both irradiations, temperature was monitored to be $600^{\circ}\mathrm{C}\pm10^{\circ}\mathrm{C}$. Heavy-ions fluxes (C, Fe, He) have been measured using Faraday cups. \\
\noindent
After irradiation, samples were polished with $0.05\mu$m colloidal silica with a vibratory polisher to remove 200nm (resp. 1$\mu m$) for C irradiation (resp. Fe irradiation) from the irradiated surface. The thickness of material removal was controlled by monitoring the evolution of micro-hardness indents sizes. Some tensile samples were subjected to uniaxial stress loading conditions at $300^{\circ}$C in a conventional tensile machine up to a conventional strain of 30\%, while others were kept to assess the initial microstructure after irradiation before mechanical loading. TEM foils were extracted at the samples' surface through conventional polishing technique \cite{gupta2016}, and observed with a FEI Tecnai G2 300kV Transmission Electron Microscope (TEM). The presence of cavities was assessed using under and over focus technique, and quantified with an in-house image analysis software. \\

\newpage
\noindent
\textbf{Numerical simulations: Polycrystalline aggregate}\\

\begin{figure}[H]
\centering
\subfigure[]{\includegraphics[height = 5cm]{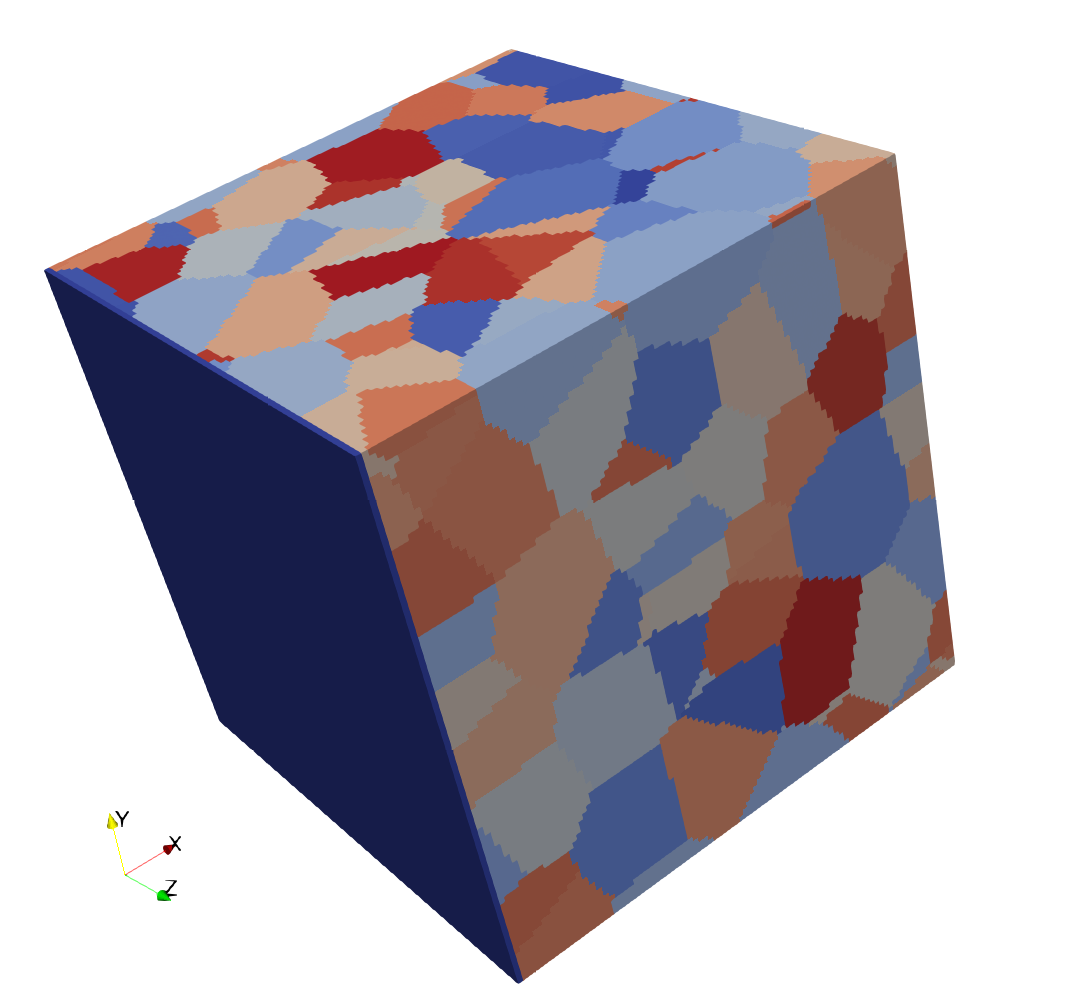}}
\subfigure[]{\includegraphics[height = 5cm]{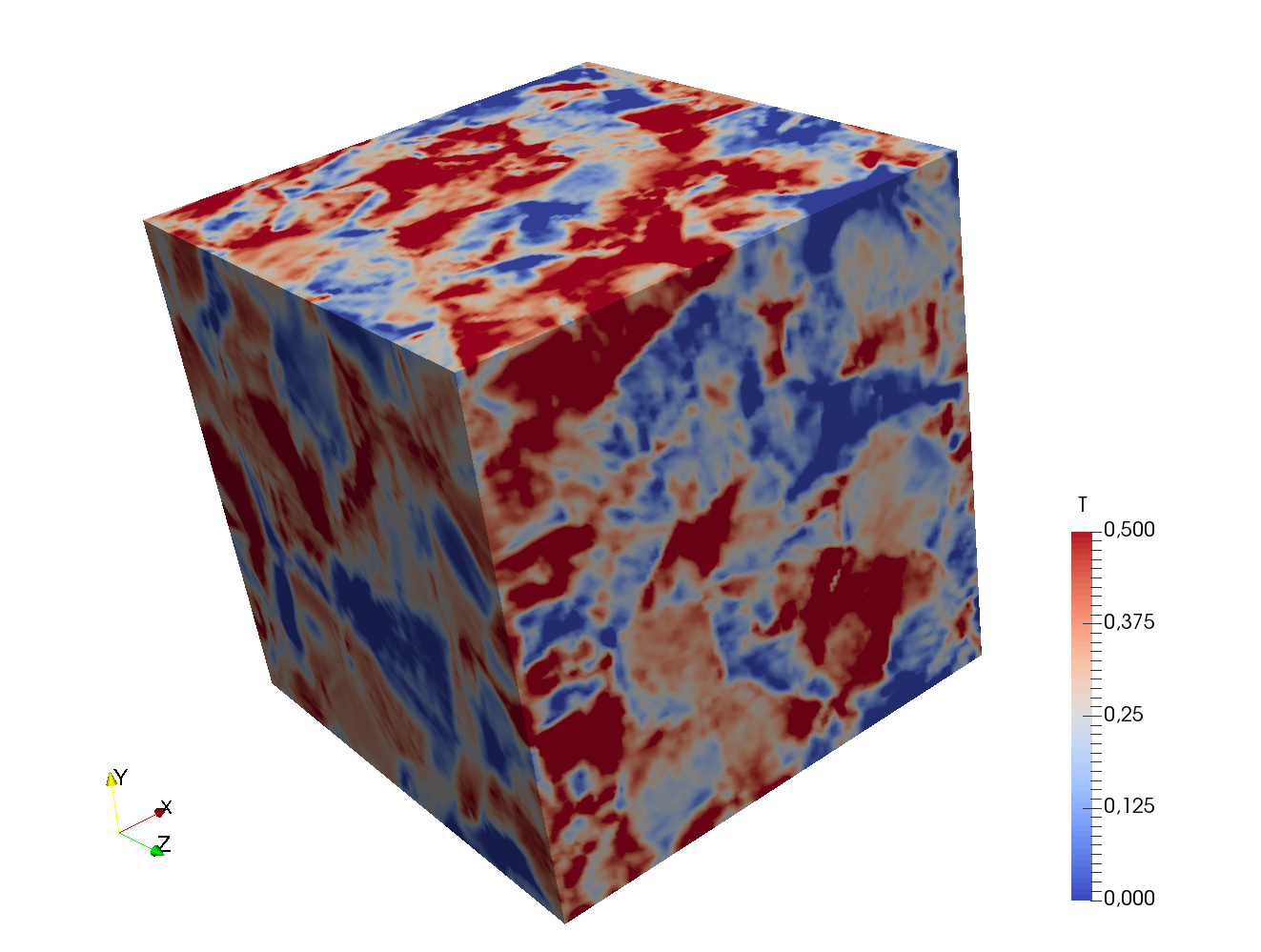}}\\
\subfigure[]{\includegraphics[height = 5cm]{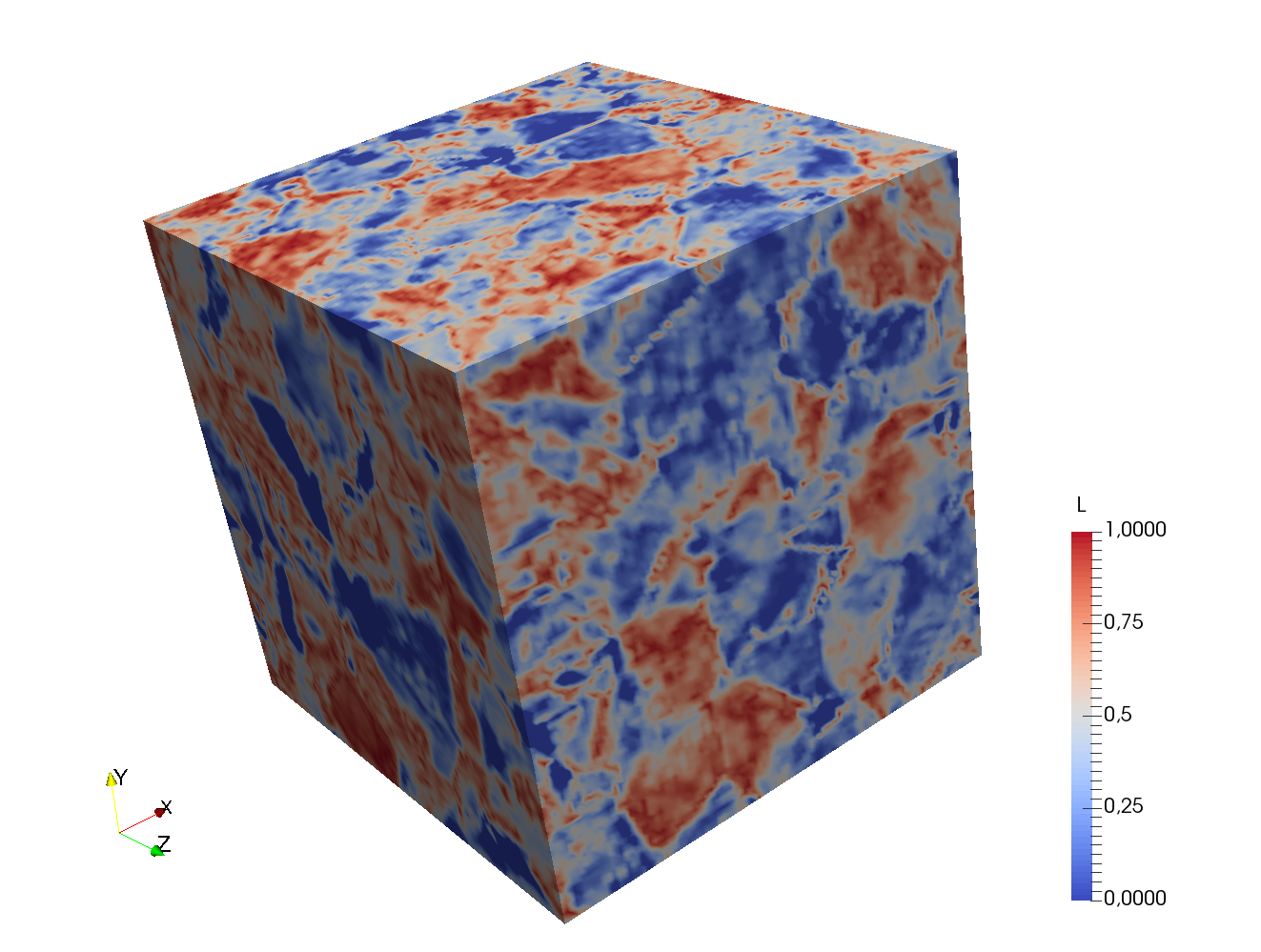}}
\subfigure[]{\includegraphics[height = 5cm]{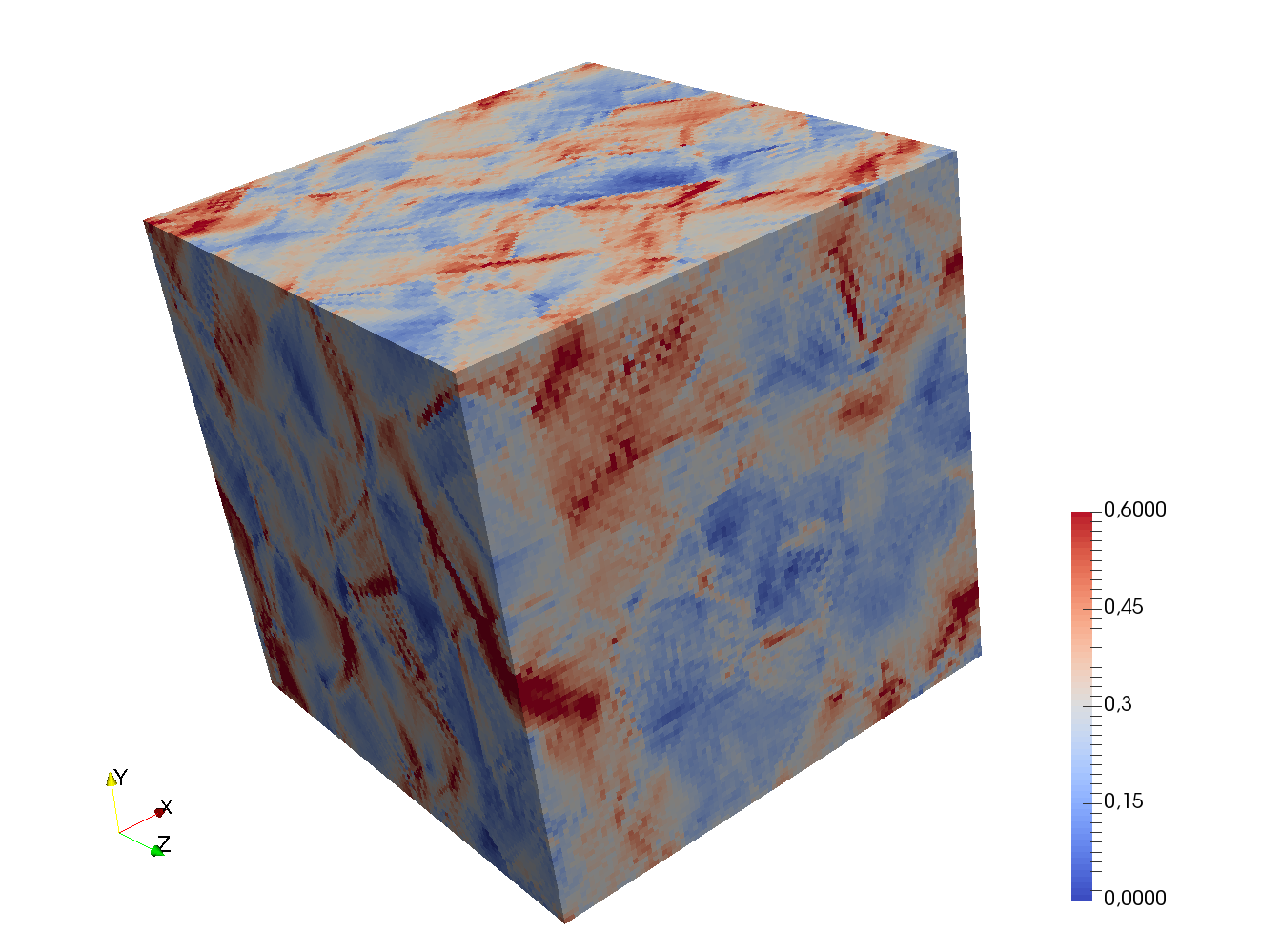}}
\caption{(a) Structured grid used to perform the simulation. Each color corresponds to different crystallographic orientations, the blue plane on the left side is the free surface models as a stress-free material. (b,c,d) Results of the simulation: local triaxiality (b), Lode parameter (c) and axial deformation gradient (d) fields.}
\label{sup2}
\end{figure}

Uniaxial stress loading conditions are applied to polycrystalline tensile samples in the experimental methodology, while Transmission Electron Microscope observations are performed at the single grain scale at the surface of the tensile samples. Numerical simulations have thus been performed to evaluate the distributions of local stresses / strains, similarly to what has been reported in \cite{barbe2}. Three-dimensionnal Voronoi aggregates are considered, each grain corresponding to a random crystallographic orientation. A free surface is considered along one direction to evaluate its effect on local stress/strain fields. Finite strain FCC crystal plasticity constitutive equations derived and calibrated for austenitic stainless steels are used. Details about the equations as well as finite strain framework, numerical implementation and parameters can be found in \cite{hure2016}. Simulations are performed with \texttt{AMITEX\_FFTP} solver which is based on the Fast Fourier Transform method \cite{amitex}. Mechanical equilibrium is solved on periodic structured grids where each voxel can be assigned constitutive equations. A typical grid used for the simulations is given in Fig.~\ref{sup2}a.

\begin{figure}[H]
\centering
\subfigure[]{\includegraphics[height = 4.5cm]{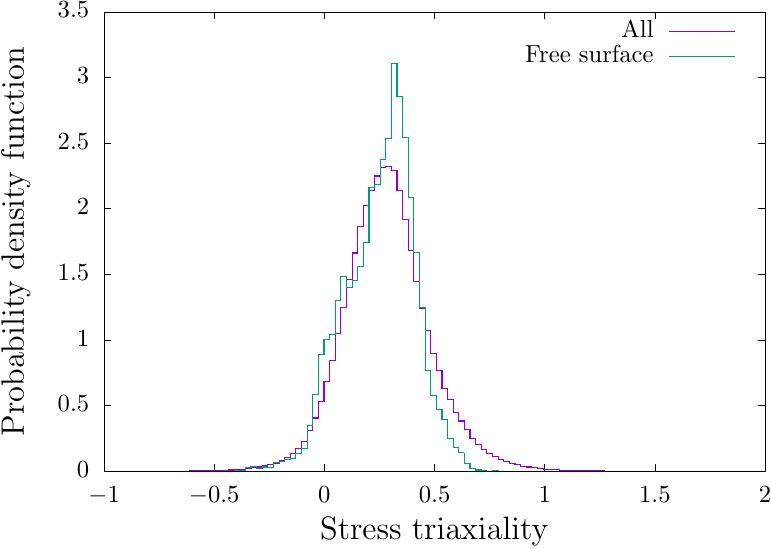}}
\hspace{1cm}
\subfigure[]{\includegraphics[height = 4.5cm]{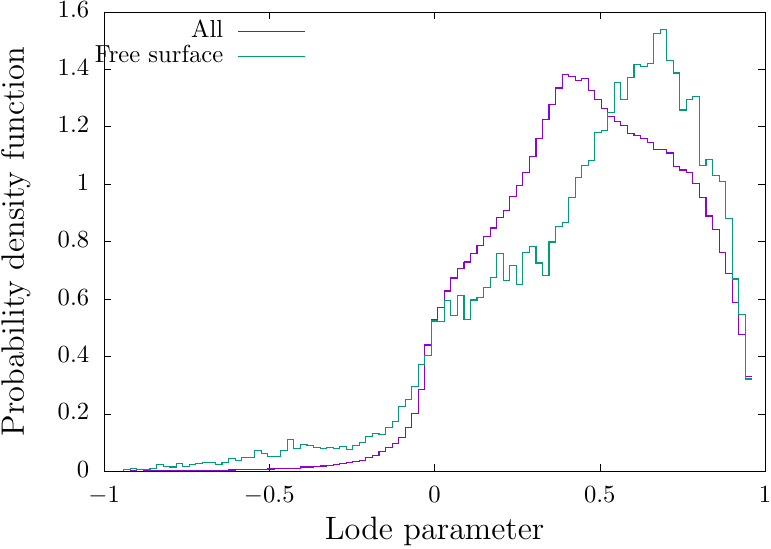}}
\caption{Probability density function of stress triaxiality (a) and Lode parameter (b) in the polycrystalline aggregate shown on Fig.~\ref{sup2}a, considering the entire aggregate or only one layer at the free surface}
\label{sup3}
\end{figure}

\newpage 
Macroscopic (volume average) uniaxial stress loading conditions are applied to the aggregate up to a macroscopic conventional strain of 30\%. Local stress fields are characterized by looking at stress triaxiality $T$ - ratio of the mean stress $\sigma_m = \sigma_{kk}/3$ over equivalent von Mises stress $\sigma_{eq} = \sqrt{[3/2]\sigma':\sigma'}$ (where $\sigma'$ is the stress deviator defined such as $\sigma' = \sigma - \sigma_m$) - and Lode parameter $L = -3[\sigma'_{II}]/[\sigma'_{I} - \sigma'_{III}]$, where $\sigma'_{I,II,III}$ are the principal components of the stress deviator in descending order \cite{defaisse}. For uniaxial stress loading conditions, triaxiality and Lode parameters are equal to $1/3$ and $1$, respectively. Local strain fields are characterized by the axial (along the tensile axis) deformation gradient $F_{zz}$, which is equal to 1.3 at the macroscopic level. Results for a 200 grains aggregate discretized with 1M voxels are described in the following, but lower number of grains (100) and coarser discretization (125k) have been checked to lead to quantitatively similar results. Significant deviations from macroscopic strain/stress uniaxial conditions appear at the grain scale, as shown in Fig.~\ref{sup2}b,c,d, and quantified in Fig.~\ref{sup3} through the estimation of probability density functions of both stress triaxiality and Lode Parameter. Close to the free surface (Fig.~\ref{sup3}), local conditions closer to uniaxial stress conditions are observed.   

\noindent
In a nutshell, strong deviations of uniaxial stress loading conditions appear at the local scale in polycrystalline aggregate subjected to tensile loading. Key points are that deviations are weaker close to the free surface (where TEM observations have been performed in this study) and that the typical scale associated with the local heterogeneities is the grain scale.\\

\noindent
\textbf{Numerical simulations: Surface tension}\\

\begin{figure}[H]
\centering
\subfigure[]{\includegraphics[height = 4.5cm]{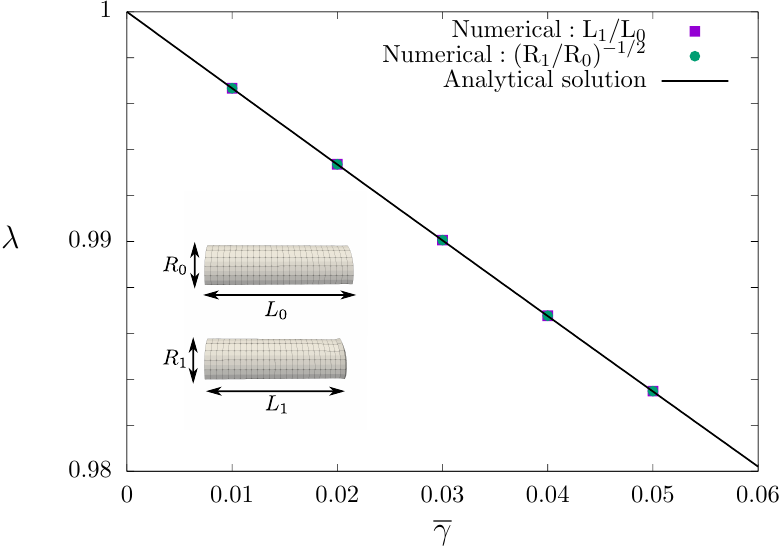}}
\hspace{0.5cm}
\subfigure[]{\includegraphics[height = 4.5cm]{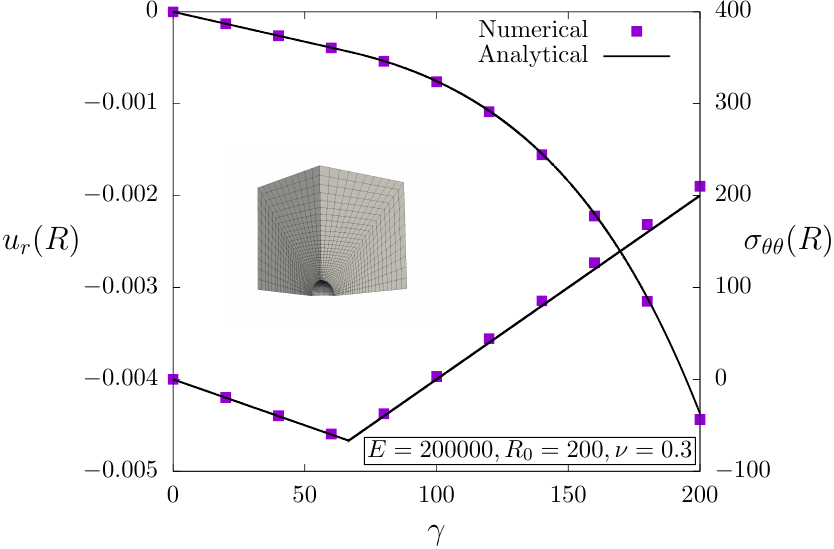}}
\caption{Comparisons between analytical solutions and numerical results for (a) an incompressible elastic cylinder accounting for surface tension on the external surface (b) a cavity of unit radius inside an elastoplastic material. Insets: Meshes used for the numerical simulations.}
\label{sup4}
\end{figure}

In order to model constant isotropic surface tension of magnitude $\gamma$ through finite element simulations, shell elements are added at the considered interface with constant isotropic 2D in-plane stresses (independent of strains), as proposed in \cite{leblondsurfacetension}:
\begin{equation}
  \textbf{N} = \left[  \begin{matrix} \gamma & 0 \\ 0 & \gamma  \end{matrix} \right]
  \label{st}
\end{equation}
All finite element simulations have been performed with the solver \texttt{Cast3M} \cite{castem}, using fully-integrated three-dimensional quadratic elements for meshing solids. Interfaces where surface tension are accounted for are meshed with two dimensional linear \texttt{DKT} shell elements, and the constitutive behavior (Eq.~\ref{st}) is introduced through a \texttt{User MATerial (UMAT)} subroutine. In order to validate this modelling of surface tension, two test cases have been performed, where numerical results have been compared to analytical predictions. The first test corresponds to an (infinite) length incompressible elastic cylinder of shear modulus $\mu$ and radius $R$ \cite{mora}. The external surface of the cylinder has a surface tension $\gamma$, leading to a shortening of the cylinder such that any axial length $L$ ($\gg R$) and the external radius $R$ become $\lambda L$ and $\lambda^{-2} R$, respectively, with:
\begin{equation}
  \lambda = \left[ \left(1 + \frac{\overline{\gamma}}{4}  - \frac{\overline{\gamma}}{2}  \right)^{1/2}   \right]^{2/3}\ \ \ \ \ \mathrm{with} \ \ \ \ \ \overline{\gamma} = \frac{\gamma}{\mu R}
\label{test1}
\end{equation}
\noindent
Analytical solution (Eq.~\ref{test1}) is compared to the numerical prediction in Fig.~\ref{sup4}a, using the numerical assumption detailed above and considering long cylinder (typically $L \geq 10R$), and a perfect agreement is obtained. The second test is a spherical cavity of radius $R$ inside an (infinite) elastoplastic solid of Young's modulus $E$, Poisson ratio $\nu$, and (constant) yield stress $\sigma_0$, where plastic flow obeys von Mises criterion. Surface tension $\gamma$ at the void/matrix interface is equivalent to a negative pressure inside the cavity of magnitude $\sigma_{rr}(R) = p = 2\gamma/R$ in spherical coordinates. The radial displacement and the stresses at the void/matrix interface are in spherical coordinates \cite{kachanov}:
\begin{equation}
  \mathrm{Elasticity}\ \left(\displaystyle{p \leq \frac{2}{3}\sigma_0}     \right)  \hspace{1cm} \left\{ \begin{matrix} \displaystyle{u_r(R) = (1+\nu) \frac{pR}{2E}}\\ \displaystyle{\sigma_{\theta \theta}(R) = \sigma_{\phi \phi}(R) = -\frac{p}{2}} \end{matrix}    \right.
\label{test2a}
\end{equation}
\begin{equation}
  \mathrm{Plasticity}\ \left(\displaystyle{p > \frac{2}{3}\sigma_0}     \right)  \hspace{1cm} \left\{ \begin{matrix} \displaystyle{u_r(R) = \frac{\sigma_0}{E}R \left((1-\nu)\frac{c^3}{R^3} - \frac{2}{3}(1-2\nu)\left( 1 + 3 \ln\left(\frac{c}{R}   \right)  \right)       \right)}\\ \displaystyle{\sigma_{\theta \theta}(R) = \sigma_{\phi \phi}(R) = \frac{2}{3}\sigma_0 \left(\frac{1}{2} - 3\ln\left(\frac{c}{R}  \right)   \right)} \end{matrix}    \right.
  \label{test2b}
\end{equation}
with $\displaystyle{c = R \exp\left( \frac{p}{2\sigma_0} - \frac{1}{3}   \right)  }$. As for the shortening of the elastic cylinder, a very good agrement is obtained between analytical solutions (Eqs.~\ref{test2a},~\ref{test2b}) and numerical results (Fig.~4b), validating the modelling of surface tension in finite element simulations.
\noindent

\newpage
Simulations reported in the manuscript have been performed with the following assumptions:
\begin{itemize}
\item[$\bullet$] Porous cubic unit-cell of length $L$ with initially spherical cavity of radius $R$, with initial porosity $\displaystyle{f = \frac{4\pi R^3}{3L^3} = 1\%}$
\item[$\bullet$] Periodic boundary conditions $\underline{u}(\underline{x}) = \bm{E}.\underline{x} + \underline{u}^{\star}(\underline{x})$, where $\bm{E}$ is the macroscopic (volume-average) strain tensor and $\underline{u}^{\star}$ a periodic fluctuation
\end{itemize}
These standard conditions (see, \textit{e.g.}, \cite{yerra}) allow to simulate accurately porous single crystals, with the assumption of a periodic cubic array of voids. The porosity used is in agreement wih the experimental observations. A typical mesh is shown in Fig.~5a.
\begin{itemize}
\item[$\bullet$] Uniaxial stress loading conditions (all stress tensor components are equal to zero, except $\sigma_{xx}$), where macroscopic axial strain $E_{xx}$ is prescribed. A strain rate of $10^{-5}\mathrm{s^{-1}}$ is used, consistently with the experiments.
\end{itemize}
Uniaxial stress loading conditions are applied at the single grain scale, keeping in mind that deviations from these conditions (applied only at the tensile sample scale) may occured at local scale, as detailed earlier.
\begin{itemize}
 \item[$\bullet$] Surface tension $\gamma$ at the void/matrix interface, and crystal plasticity constitutive equations for the matrix \cite{hure2016}
\end{itemize}
Surface tension is assumed to be constant and independent of the interface orientation. The parameters of the crystal plasticity constitutive equations should be calibrated for the irradiated matrix material. However, in absence of data, parameters determined for unirradiated 304L stainless steel are used. Different random crystallographic orientations are used.\\

\noindent
For each simulation, void shape is monitored (Fig.~5b) and the aspect ratio is measured as follows: void shape is fitted by an ellipsoid, with $l_1 \geq l_2 \geq l_3$ the axis length, and the mean aspect ratio $a_r$ and mean void diameter $D$ are defined as $a_r = \sqrt{l_1/l_2}\sqrt{l_1/l_3} $ and $D = \sqrt[3]{l_1 l_2 l_3}$, in order to be close to the experimental conditions.  

\begin{figure}[H]
\centering
\subfigure[]{\includegraphics[height = 4.5cm]{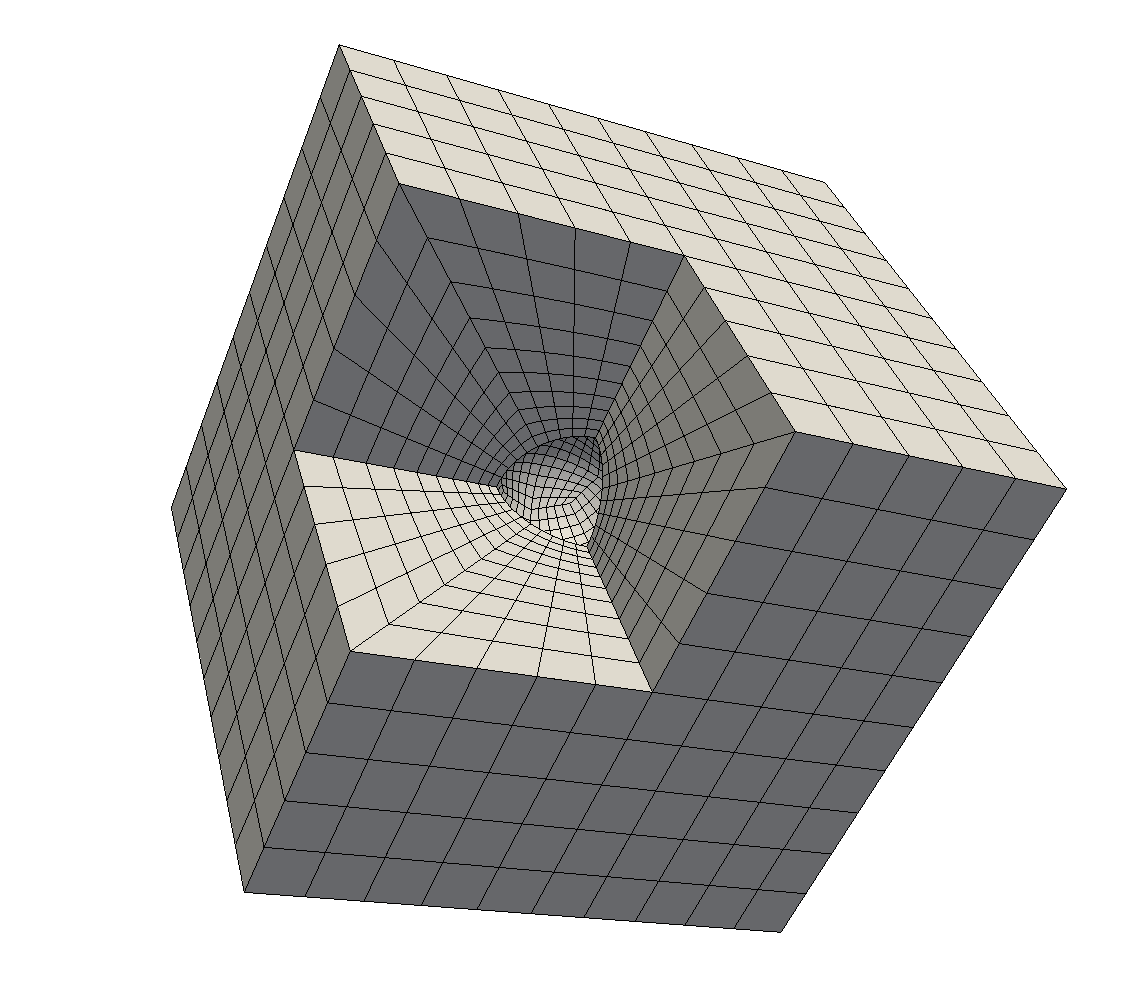}}
\subfigure[]{\includegraphics[height = 4.5cm]{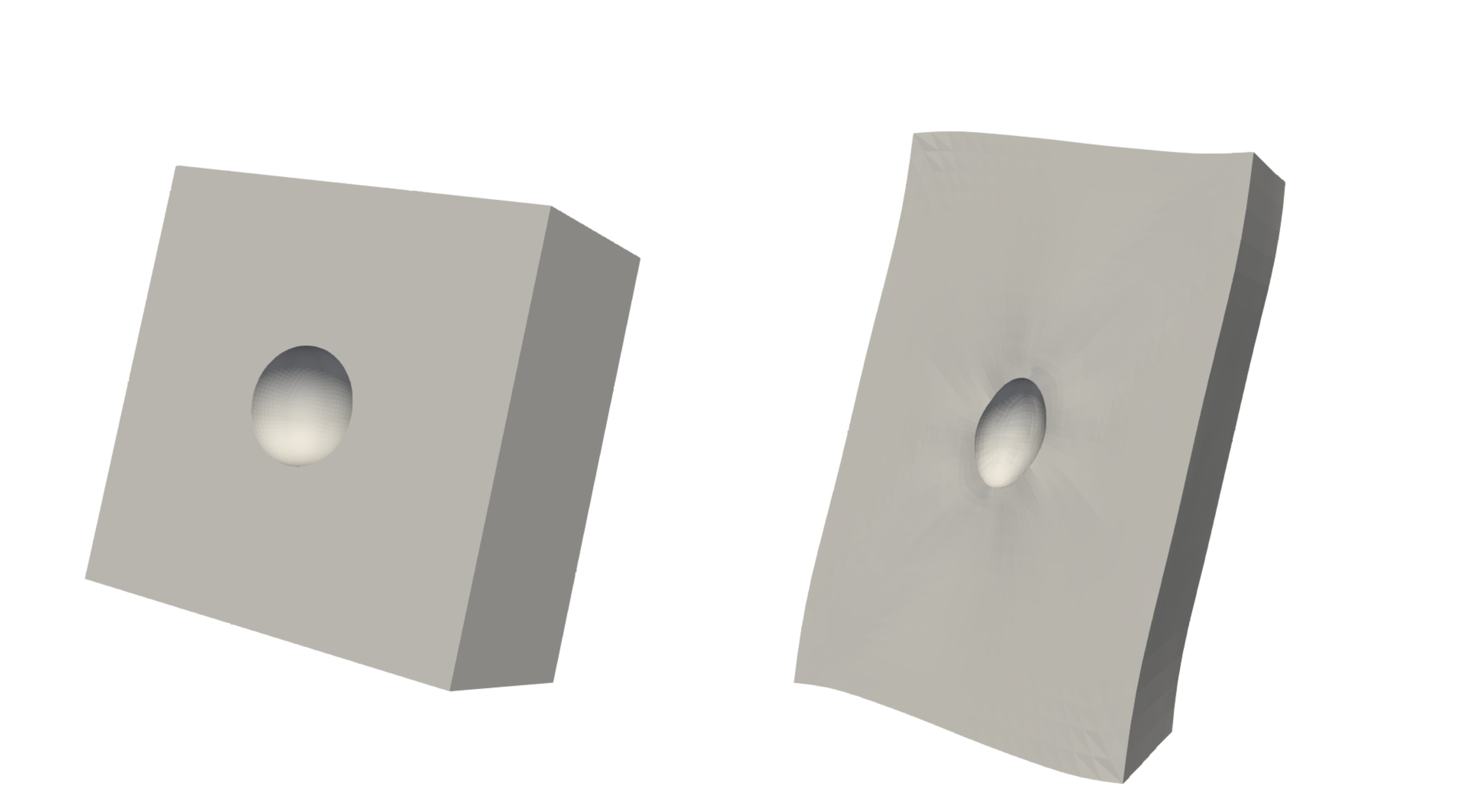}}
\caption{(a) Typical mesh used for the simulations, where one-eight of the elements have been removed to allow visualizating the cavity (b) Typical void shape evolution after straining up to 30\%.}
\label{sup5}
\end{figure}

\bibliographystyle{model1a-num-names.bst}
\bibliography{spebib2}

\begin{thebibliography}{49}
\expandafter\ifx\csname natexlab\endcsname\relax\def\natexlab#1{#1}\fi
\providecommand{\url}[1]{\texttt{#1}}
\providecommand{\href}[2]{#2}
\providecommand{\path}[1]{#1}
\providecommand{\DOIprefix}{doi:}
\providecommand{\ArXivprefix}{arXiv:}
\providecommand{\URLprefix}{URL: }
\providecommand{\Pubmedprefix}{pmid:}
\providecommand{\doi}[1]{\href{http://dx.doi.org/#1}{\path{#1}}}
\providecommand{\Pubmed}[1]{\href{pmid:#1}{\path{#1}}}
\providecommand{\bibinfo}[2]{#2}
\ifx\xfnm\relax \def\xfnm[#1]{\unskip,\space#1}\fi
\bibitem[{Hall(1951)}]{hall}
\bibinfo{author}{E.~Hall}, \bibinfo{journal}{Proc. Phys. Soc. B}
  \bibinfo{volume}{61} (\bibinfo{year}{1951}) \bibinfo{pages}{747--753}.
\bibitem[{Petch(1953)}]{petch}
\bibinfo{author}{N.~Petch}, \bibinfo{journal}{J. Iron Steel Institute}
  \bibinfo{volume}{174} (\bibinfo{year}{1953}) \bibinfo{pages}{25--28}.
\bibitem[{Gane and Cox(1970)}]{gane}
\bibinfo{author}{M.~Gane}, \bibinfo{author}{J.~Cox}, \bibinfo{journal}{Phil.
  Mag.} \bibinfo{volume}{22} (\bibinfo{year}{1970}) \bibinfo{pages}{881--891}.
\bibitem[{Stolken and Evans(1998)}]{stolken}
\bibinfo{author}{J.~Stolken}, \bibinfo{author}{A.~Evans},
  \bibinfo{journal}{Acta Mater.} \bibinfo{volume}{46} (\bibinfo{year}{1998})
  \bibinfo{pages}{5109--5115}.
\bibitem[{Ashby(1970)}]{ashby}
\bibinfo{author}{M.~Ashby}, \bibinfo{journal}{Phil. Mag.} \bibinfo{volume}{21}
  (\bibinfo{year}{1970}) \bibinfo{pages}{399--424}.
\bibitem[{Smyshlyaev and Fleck(1996)}]{smyshlyaev}
\bibinfo{author}{V.~Smyshlyaev}, \bibinfo{author}{N.~Fleck},
  \bibinfo{journal}{J. Mech. Phys. Solids} \bibinfo{volume}{44}
  (\bibinfo{year}{1996}) \bibinfo{pages}{465--495}.
\bibitem[{Nix and Gao(1998)}]{nixgao}
\bibinfo{author}{W.~Nix}, \bibinfo{author}{H.~Gao}, \bibinfo{journal}{J. Mech.
  Phys. Solids} \bibinfo{volume}{46} (\bibinfo{year}{1998})
  \bibinfo{pages}{411--425}.
\bibitem[{Lilleodden and Nix(2006)}]{lilleodden}
\bibinfo{author}{E.~Lilleodden}, \bibinfo{author}{W.~Nix},
  \bibinfo{journal}{Acta Mater.} \bibinfo{volume}{54} (\bibinfo{year}{2006})
  \bibinfo{pages}{1583--1593}.
\bibitem[{Kramer et~al.(2016)Kramer, Viswanath, and Weissmuller}]{kramer}
\bibinfo{author}{D.~Kramer}, \bibinfo{author}{R.~Viswanath},
  \bibinfo{author}{J.~Weissmuller}, \bibinfo{journal}{Nano Let.}
  \bibinfo{volume}{4} (\bibinfo{year}{2016}) \bibinfo{pages}{793--796}.
\bibitem[{Fleck and Hutchinson(1994)}]{fleckhutchinson}
\bibinfo{author}{N.~Fleck}, \bibinfo{author}{J.~Hutchinson},
  \bibinfo{journal}{J. Mech. Phys. Solids} \bibinfo{volume}{41}
  (\bibinfo{year}{1994}) \bibinfo{pages}{1825--1857}.
\bibitem[{Fleck and Hutchinson(1997)}]{fleckhutchinson2}
\bibinfo{author}{N.~Fleck}, \bibinfo{author}{J.~Hutchinson},
  \bibinfo{journal}{Adv. Appl. Mech.} \bibinfo{volume}{33}
  (\bibinfo{year}{1997}) \bibinfo{pages}{295--361}.
\bibitem[{Forest(2009)}]{forest}
\bibinfo{author}{S.~Forest}, \bibinfo{journal}{J. Eng. Mech.}
  \bibinfo{volume}{135} (\bibinfo{year}{2009}) \bibinfo{pages}{117--131}.
\bibitem[{Benzerga and Leblond(2010)}]{benzergaleblond}
\bibinfo{author}{A.~A. Benzerga}, \bibinfo{author}{J.-B. Leblond},
  \bibinfo{journal}{Adv. Applied Mech.} \bibinfo{volume}{44}
  (\bibinfo{year}{2010}) \bibinfo{pages}{169--305}.
\bibitem[{Traiviratana et~al.(2008)Traiviratana, Bringa, Benson, and
  Meyers}]{traiviratana}
\bibinfo{author}{S.~Traiviratana}, \bibinfo{author}{E.~Bringa},
  \bibinfo{author}{D.~Benson}, \bibinfo{author}{M.~Meyers},
  \bibinfo{journal}{Acta Mat.} \bibinfo{volume}{56} (\bibinfo{year}{2008})
  \bibinfo{pages}{3874--3886}.
\bibitem[{Chang et~al.(2013)Chang, Segurado, Rodriguez de~la Fuente, Pabon, and
  Llorca}]{chang2013}
\bibinfo{author}{H.~Chang}, \bibinfo{author}{J.~Segurado},
  \bibinfo{author}{O.~Rodriguez de~la Fuente}, \bibinfo{author}{B.~Pabon},
  \bibinfo{author}{J.~Llorca}, \bibinfo{journal}{Modelling Simul. Mater. Sci.
  Eng.} \bibinfo{volume}{21} (\bibinfo{year}{2013}) \bibinfo{pages}{075010}.
\bibitem[{Segurado and Llorca(2009)}]{segurado2009}
\bibinfo{author}{J.~Segurado}, \bibinfo{author}{J.~Llorca},
  \bibinfo{journal}{Acta Mat.} \bibinfo{volume}{57} (\bibinfo{year}{2009})
  \bibinfo{pages}{1427--1436}.
\bibitem[{Chang et~al.(2015)Chang, Segurado, and Llorca}]{chang2015}
\bibinfo{author}{H.~Chang}, \bibinfo{author}{J.~Segurado},
  \bibinfo{author}{J.~Llorca}, \bibinfo{journal}{Scripta Mat.}
  \bibinfo{volume}{95} (\bibinfo{year}{2015}) \bibinfo{pages}{11--14}.
\bibitem[{Borg et~al.(2008)Borg, Niordson, and Kysar}]{borg2008}
\bibinfo{author}{U.~Borg}, \bibinfo{author}{C.~Niordson},
  \bibinfo{author}{J.~Kysar}, \bibinfo{journal}{Int. J. Plasticity}
  \bibinfo{volume}{24} (\bibinfo{year}{2008}) \bibinfo{pages}{688--701}.
\bibitem[{Wen et~al.(2005)Wen, Huang, Hwang, Kiu, and Li}]{wen2005}
\bibinfo{author}{J.~Wen}, \bibinfo{author}{Y.~Huang},
  \bibinfo{author}{K.~Hwang}, \bibinfo{author}{C.~Kiu},
  \bibinfo{author}{M.~Li}, \bibinfo{journal}{Int. J. Plasticity}
  \bibinfo{volume}{21} (\bibinfo{year}{2005}) \bibinfo{pages}{381--395}.
\bibitem[{Monchiet and Bonnet(2013)}]{monchiet2013b}
\bibinfo{author}{V.~Monchiet}, \bibinfo{author}{G.~Bonnet},
  \bibinfo{journal}{Int. J. Sol. Struc.} \bibinfo{volume}{50}
  (\bibinfo{year}{2013}) \bibinfo{pages}{320--327}.
\bibitem[{Niordson(2008)}]{niordson2008}
\bibinfo{author}{C.~Niordson}, \bibinfo{journal}{Eur. J. Mech. A/Solids}
  \bibinfo{volume}{27} (\bibinfo{year}{2008}) \bibinfo{pages}{222--233}.
\bibitem[{Ziegler et~al.(2010)Ziegler, Ziegler, and Biersack}]{SRIM}
\bibinfo{author}{J.~Ziegler}, \bibinfo{author}{M.~Ziegler},
  \bibinfo{author}{J.~Biersack}, \bibinfo{journal}{Nuc. Ins. Methods in Phys.
  B} \bibinfo{volume}{268} (\bibinfo{year}{2010}) \bibinfo{pages}{1818--1823}.
\bibitem[{Margolin et~al.(2016)Margolin, Minkin, Smirnov, Sorokin, Shvetsova,
  and Potapova}]{margolin2016}
\bibinfo{author}{B.~Margolin}, \bibinfo{author}{A.~Minkin},
  \bibinfo{author}{V.~Smirnov}, \bibinfo{author}{A.~Sorokin},
  \bibinfo{author}{V.~Shvetsova}, \bibinfo{author}{V.~Potapova},
  \bibinfo{journal}{J. Nuc. Mat.} \bibinfo{volume}{480} (\bibinfo{year}{2016})
  \bibinfo{pages}{52--68}.
\bibitem[{Neustroev and Garner(2008)}]{neustroev}
\bibinfo{author}{V.~Neustroev}, \bibinfo{author}{F.~Garner},
  \bibinfo{journal}{J. Nuc. Mat.} \bibinfo{volume}{378} (\bibinfo{year}{2008})
  \bibinfo{pages}{327--332}.
\bibitem[{Ding et~al.(2016)Ding, Du, Wan, Ogata, Tian, Ma, Han, Li, and
  Shan}]{ding}
\bibinfo{author}{M.~Ding}, \bibinfo{author}{J.~Du}, \bibinfo{author}{L.~Wan},
  \bibinfo{author}{S.~Ogata}, \bibinfo{author}{L.~Tian},
  \bibinfo{author}{E.~Ma}, \bibinfo{author}{W.~Han}, \bibinfo{author}{J.~Li},
  \bibinfo{author}{Z.~Shan}, \bibinfo{journal}{Nano Let.} \bibinfo{volume}{16}
  (\bibinfo{year}{2016}) \bibinfo{pages}{4118--4124}.
\bibitem[{Han et~al.(2018)Han, Ding, and Shan}]{han}
\bibinfo{author}{W.~Han}, \bibinfo{author}{M.~Ding}, \bibinfo{author}{Z.~Shan},
  \bibinfo{journal}{Scripta Mat.} \bibinfo{volume}{147} (\bibinfo{year}{2018})
  \bibinfo{pages}{1--5}.
\bibitem[{Pellegrino et~al.(2012)Pellegrino, Trocellier, Miro, Serruys, Bordas,
  Martin, Cha{\^a}bane, Vaubaillon, Gallien, and Beck}]{jannus}
\bibinfo{author}{S.~Pellegrino}, \bibinfo{author}{P.~Trocellier},
  \bibinfo{author}{S.~Miro}, \bibinfo{author}{Y.~Serruys},
  \bibinfo{author}{E.~Bordas}, \bibinfo{author}{H.~Martin},
  \bibinfo{author}{N.~Cha{\^a}bane}, \bibinfo{author}{S.~Vaubaillon},
  \bibinfo{author}{J.~Gallien}, \bibinfo{author}{L.~Beck},
  \bibinfo{journal}{Nuc. Ins. Methods in Phys. B} \bibinfo{volume}{273}
  (\bibinfo{year}{2012}) \bibinfo{pages}{213--217}.
\bibitem[{Cawthorne and Fulton(1967)}]{cawthorne}
\bibinfo{author}{C.~Cawthorne}, \bibinfo{author}{E.~Fulton},
  \bibinfo{journal}{Nature} \bibinfo{volume}{216} (\bibinfo{year}{1967})
  \bibinfo{pages}{575--576}.
\bibitem[{Wulff(1901)}]{wulff}
\bibinfo{author}{G.~Wulff}, \bibinfo{journal}{Z. Krist.} \bibinfo{volume}{34}
  (\bibinfo{year}{1901}) \bibinfo{pages}{449--530}.
\bibitem[{Was(2007)}]{was}
\bibinfo{author}{G.~Was}, \bibinfo{title}{Fundamentals of {R}adiation
  {M}aterials {S}cience}, \bibinfo{publisher}{Springer-{V}erlag},
  \bibinfo{year}{2007}.
\bibitem[{{Materials Virtual Lab}(2013)}]{crystallium2}
\bibinfo{author}{{Materials Virtual Lab}}, \bibinfo{title}{Crystalium},
  \bibinfo{year}{2013}.
  \bibinfo{note}{{http://crystalium.materialsvirtuallab.org/}}.
\bibitem[{M{\'e}ric~de Bellefon and van Duysen(2016)}]{bellefon}
\bibinfo{author}{G.~M{\'e}ric~de Bellefon}, \bibinfo{author}{J.~van Duysen},
  \bibinfo{journal}{J. Nucl. Mat.} \bibinfo{volume}{475} (\bibinfo{year}{2016})
  \bibinfo{pages}{168--191}.
\bibitem[{Weck et~al.(2008)Weck, Wilkinson, Maire, and Toda}]{wecktoda}
\bibinfo{author}{A.~Weck}, \bibinfo{author}{D.~Wilkinson},
  \bibinfo{author}{E.~Maire}, \bibinfo{author}{H.~Toda}, \bibinfo{journal}{Acta
  Mat.} \bibinfo{volume}{56} (\bibinfo{year}{2008})
  \bibinfo{pages}{2919--2928}.
\bibitem[{Lecarme et~al.(2014)Lecarme, Maire, Kumar, De~Vleeschouwer, Jacques,
  Simar, and Pardoen}]{lecarmemaire}
\bibinfo{author}{L.~Lecarme}, \bibinfo{author}{E.~Maire},
  \bibinfo{author}{A.~Kumar}, \bibinfo{author}{C.~De~Vleeschouwer},
  \bibinfo{author}{L.~Jacques}, \bibinfo{author}{A.~Simar},
  \bibinfo{author}{T.~Pardoen}, \bibinfo{journal}{Acta Mater.}
  \bibinfo{volume}{63} (\bibinfo{year}{2014}) \bibinfo{pages}{130--139}.
\bibitem[{Yerra et~al.(2010)Yerra, Tekoglu, Scheyvaerts, Delannay, Van~Houtte,
  and Pardoen}]{yerra}
\bibinfo{author}{S.~Yerra}, \bibinfo{author}{C.~Tekoglu},
  \bibinfo{author}{F.~Scheyvaerts}, \bibinfo{author}{L.~Delannay},
  \bibinfo{author}{P.~Van~Houtte}, \bibinfo{author}{T.~Pardoen},
  \bibinfo{journal}{Int. J. Solids and Structures} \bibinfo{volume}{47}
  (\bibinfo{year}{2010}) \bibinfo{pages}{1016--1029}.
\bibitem[{Paccou et~al.(2019)Paccou, Tanguy, and Legros}]{paccou}
\bibinfo{author}{E.~Paccou}, \bibinfo{author}{B.~Tanguy},
  \bibinfo{author}{M.~Legros}, \bibinfo{journal}{Scripta Mat.}
  \bibinfo{volume}{172} (\bibinfo{year}{2019}) \bibinfo{pages}{56--60}.
\bibitem[{Fran{\c c}ois et~al.(2012)Fran{\c c}ois, Pineau, and Zaoui}]{fpz}
\bibinfo{author}{D.~Fran{\c c}ois}, \bibinfo{author}{A.~Pineau},
  \bibinfo{author}{A.~Zaoui}, \bibinfo{title}{Mechanical {B}ehavior of
  {M}aterials: {V}olume 1: {E}lasticity and {P}lasticity},
  \bibinfo{publisher}{Kluwer {A}cademic {P}ublishers}, \bibinfo{year}{2012}.
\bibitem[{Rice and Tracey(1969)}]{ricetracey}
\bibinfo{author}{J.~R. Rice}, \bibinfo{author}{D.~M. Tracey},
  \bibinfo{journal}{J. Mech. Phys. Solids} \bibinfo{volume}{17}
  (\bibinfo{year}{1969}) \bibinfo{pages}{201--217}.
\bibitem[{Huang(1991)}]{huang}
\bibinfo{author}{Y.~Huang}, \bibinfo{journal}{J. Appl. Mech.}
  \bibinfo{volume}{58} (\bibinfo{year}{1991}) \bibinfo{pages}{1084--1086}.
\bibitem[{Dormieux and Kondo(2010)}]{dormieux2010}
\bibinfo{author}{L.~Dormieux}, \bibinfo{author}{D.~Kondo},
  \bibinfo{journal}{Int. J. Eng. Sci.} \bibinfo{volume}{48}
  (\bibinfo{year}{2010}) \bibinfo{pages}{575--581}.
\bibitem[{Gupta et~al.(2016)Gupta, Hure, Tanguy, Laffont, Lafont, and
  Andrieu}]{gupta2016}
\bibinfo{author}{J.~Gupta}, \bibinfo{author}{J.~Hure},
  \bibinfo{author}{B.~Tanguy}, \bibinfo{author}{L.~Laffont},
  \bibinfo{author}{M.~Lafont}, \bibinfo{author}{E.~Andrieu},
  \bibinfo{journal}{J. Nuc. Mat.} \bibinfo{volume}{476} (\bibinfo{year}{2016})
  \bibinfo{pages}{82--92}.
\bibitem[{Barbe et~al.(2001)Barbe, Forest, and Cailletaud}]{barbe2}
\bibinfo{author}{F.~Barbe}, \bibinfo{author}{S.~Forest},
  \bibinfo{author}{G.~Cailletaud}, \bibinfo{journal}{Int. J. Plasticity}
  \bibinfo{volume}{17} (\bibinfo{year}{2001}) \bibinfo{pages}{537--563}.
\bibitem[{Hure et~al.(2016)Hure, El~Shawish, Cizelj, and Tanguy}]{hure2016}
\bibinfo{author}{J.~Hure}, \bibinfo{author}{S.~El~Shawish},
  \bibinfo{author}{L.~Cizelj}, \bibinfo{author}{B.~Tanguy},
  \bibinfo{journal}{J. Nuc. Mat.} \bibinfo{volume}{476} (\bibinfo{year}{2016})
  \bibinfo{pages}{231--242}.
\bibitem[{CEA(2018)}]{amitex}
\bibinfo{author}{CEA},
  \bibinfo{title}{{http://www.maisondelasimulation.fr/projects/amitex/html/}},
  \bibinfo{year}{2018}.
\bibitem[{Defaisse et~al.(2018)Defaisse, Mazi{\`e}re, Marcin, and
  Besson}]{defaisse}
\bibinfo{author}{C.~Defaisse}, \bibinfo{author}{M.~Mazi{\`e}re},
  \bibinfo{author}{L.~Marcin}, \bibinfo{author}{J.~Besson},
  \bibinfo{journal}{Eng. Fract. Mech.} \bibinfo{volume}{194}
  (\bibinfo{year}{2018}) \bibinfo{pages}{301--318}.
\bibitem[{Leblond et~al.(2013)Leblond, El~Sayed, and
  Bergheau}]{leblondsurfacetension}
\bibinfo{author}{J.~Leblond}, \bibinfo{author}{H.~El~Sayed},
  \bibinfo{author}{J.~Bergheau}, \bibinfo{journal}{Comptes Rendus
  M{\'e}canique} \bibinfo{volume}{341} (\bibinfo{year}{2013})
  \bibinfo{pages}{770--775}.
\bibitem[{CEA(2018)}]{castem}
\bibinfo{author}{CEA}, \bibinfo{title}{Cast3M}, \bibinfo{year}{2018}.
  \bibinfo{note}{{www-cast3m.com}}.
\bibitem[{Mora et~al.(2013)Mora, Maurini, Phou, Fromental, Audoly, and
  Pomeau}]{mora}
\bibinfo{author}{S.~Mora}, \bibinfo{author}{C.~Maurini},
  \bibinfo{author}{T.~Phou}, \bibinfo{author}{J.~Fromental},
  \bibinfo{author}{B.~Audoly}, \bibinfo{author}{Y.~Pomeau},
  \bibinfo{journal}{Phys. Rev. Lett.} \bibinfo{volume}{111}
  (\bibinfo{year}{2013}) \bibinfo{pages}{114301}.
\bibitem[{Kachanov(2004)}]{kachanov}
\bibinfo{author}{L.~Kachanov}, \bibinfo{title}{Fundamentals of the theory of
  plasticity}, \bibinfo{publisher}{Dover}, \bibinfo{year}{2004}.

\end{thebibliography}

\end{document}